\documentclass[showpacs,amsmath,amssymb,aps]{revtex4}

\usepackage{graphicx}       
\usepackage{dcolumn}        
\usepackage{bm}             
\usepackage{psfrag}

\newcommand{\nn}{\nonumber}

\newcommand{\lam}{\lambda}

\newcommand{\gam}{\gamma}

\newcommand{\alp}{\alpha}

\newcommand{\sig}{\sigma}

\newcommand{\om}{\omega}
\newcommand{\eps}{\epsilon}

\newcommand{\coshalf}{\cos(\theta / 2)}
\newcommand{\sinhalf}{\sin(\theta / 2)}

\newcommand{\Sp}{S_+}
\newcommand{\Sm}{S_-}
\newcommand{\Tp}{T_+}
\newcommand{\Tm}{T_-}
\newcommand{\Par}{\mathcal{P}}

\newcommand{\eig}{ \lambda_{jm\Par}^{(a\omega, a\mu)}}

\begin{document}


 \title{The massive Dirac field on a rotating black hole spacetime: Angular solutions}

\author{Sam R. Dolan}
 \email{sam.dolan@ucd.ie}
 \affiliation{%
 School of Mathematical Sciences, University College Dublin, Belfield, Dublin 4, Ireland. \\
}%

\author{Jonathan R. Gair}
 \email{jgair@ast.cam.ac.uk}
 \affiliation{%
 Institute of Astronomy, University of Cambridge, Madingley Road, Cambridge, CB3 0HA, UK.  \\
}%

\date{\today}

\begin{abstract}

The massive Dirac equation on a Kerr-Newman background may be solved by the method of separation of variables. The radial and angular equations are coupled via an angular eigenvalue, which is determined from the Chandrasekhar-Page (CP) equation. Obtaining accurate angular eigenvalues is a key step in studying scattering, absorption and emission of the fermionic field. 

Here we introduce a new method for finding solutions of the CP equation. First, we introduce a novel representation for the spin-half spherical harmonics. Next, we decompose the angular solutions of the CP equation (the mass-dependent spin-half spheroidal harmonics) in the spherical basis. The method yields a three-term recurrence relation which may be solved numerically via continued-fraction methods, or perturbatively to obtain a series expansion for the eigenvalues. In the case $\mu = \pm \omega$ (where $\omega$ and $\mu$ are the frequency and mass of the fermion) we obtain eigenvalues and eigenfunctions in closed form. We study the eigenvalue spectrum, and the zeros of the maximally co-rotating mode. 

We compare our results with previous studies, and uncover and correct some errors in the literature. We provide series expansions, tables of eigenvalues and numerical fits across a wide parameter range, and present plots of a selection of eigenfunctions. It is hoped this study will be a useful resource for all researchers interested in the Dirac equation on a rotating black hole background.
\end{abstract}

\pacs{04.70.-s, 04.62.+v}
\maketitle

\section{\label{sec:introduction}Introduction}

The interaction of a fermionic (Dirac) field with a rotating charged (Kerr-Newman) black hole is of deep theoretical interest. The interaction depends on three couplings: coupling between quantum-mechanical spin and black hole (BH) rotation; coupling between field mass and BH mass; and coupling between field charge and BH charge. Desire for a deeper understanding of these couplings has motivated a range of studies over the last three decades.

Over thirty years ago, Chandrasekhar \cite{Chandrasekhar-1976} and Page \cite{Page-1976} showed that separation of variables is possible for the massive Dirac field on the Kerr-Newman spacetime. 
In other words, the partial differential equations (PDEs) governing the evolution of the Dirac field may be reduced to a set of coupled ordinary differential equations (ODEs), and the solution can be expressed as a sum over modes and integral over frequency. The radial and angular ODEs are coupled through an angular eigenvalue $\lambda$. The eigenvalue is found by solving the so-called \emph{Chandrasekhar-Page} (CP) equation (see Eq.~(\ref{chandrasekhar-page-eqn}) below). The eigenfunctions of the CP equation -- the so-called \emph{mass dependent spin-half spheroidal harmonics} (MDSHs) -- are required for a full reconstruction of the field. In this paper we present a new method for computing both the eigenvalues and eigenfunctions. Our aim is to show that finding MDSHs is no more difficult than finding the spin-weighted spheroidal harmonics for massless fields \cite{Berti-2005}. Along the way, we review alternative methods and highlight some  inaccuracies in the literature.

Motivation for this work arose from two separate studies, conducted independently by the present authors. The Kerr-Newman solution takes a very beautiful form in the limit where the electromagnetic field is assumed to dominate over gravity, which is obtained by letting $G\rightarrow 0$ in the metric. The result is flat-space, in oblate-spheroidal coordinates, with an electromagnetic field
\begin{equation}
{\bf E}+{\rm i}{\bf B} = - \nabla \left( \frac{1}{\sqrt{({\bf r} - {\rm i}{\bf a})\cdot ({\bf r} - {\rm i}{\bf a})}} \right)
\label{magicfield}
\end{equation}
i.e., a Coulomb field centred at the imaginary point ${\bf r} = {\rm i}{\bf a} = {\rm i} (0,0,a)$~\cite{dlb03,dlb04a,dlb04b}. This field can be used to model a rapidly rotating but effectively massless nucleus~\cite{Pekeris-1989-hyperfine} and in his thesis~\cite{Gair-thesis} JG examined the possibility that the hyperfine splitting observed in muonium could be reproduced by the coupling between electron spin and the frame dragging caused by nuclear rotation in this model~\cite{Pekeris-1989-hyperfine}. SD studied the spectrum of fermionic quasi-bound states in the vicinity of a small non-extremal black hole with $a < M$, extending recent work in this area \cite{Lasenby-2005, Dolan-2007}. Both studies required the determination of accurate eigenvalues for the angular eigen-equation, and both studies were impeded by a lack of reliable numerical results in the literature. These studies will be presented in full elsewhere. 

The massive Dirac-Kerr-Newman system has been studied in many other contexts. For example, authors have investigated the absence of fermionic superradiance \cite{Iyer-1978, Wagh-1985}; scattering and absorption  \cite{Kalnins-1992, Chakrabarti-2000, Mukhopadhyay-2000a, Mukhopadhyay-2000b, Batic-2007b}; and  Hawking radiation emission \cite{Page-1976b, Zhou-2008}. 
Our intention here is to provide a solid foundation for possible further work. Let us give three examples of future applications. First, it has been shown \cite{Finster-2002, Finster-2003, He-2006a, He-2006b, Finster-2008} that a rotating BH is stable to fermionic perturbations. Nevertheless, the quantative effect of field mass \cite{Simone-Will-1992} on the fermionic quasi-normal mode spectrum \cite{Jing-2005} has not been studied. Secondly, field mass is usually neglected when the Hawking emission process is considered \cite{Page-1976}. The emission spectrum will change significantly if the Hawking temperature approaches the mass of a  particle species of the standard model. Emission of massive scalars by a Kerr-Newman black hole was considered in \cite{Grain-Barrau-2008}, but a corresponding study for massive fermions is still lacking. Finally, in recent years there has been much interest in theories with extra dimensions which lead to the possibility of BH creation in hadron colliders \cite{Kanti-2004}. The radial equations for fermionic fields on the brane are closely related to their 4D counterparts \cite{Casals-2007}, and the angular equations are unchanged. The effect of field mass on emission of fermions from non-rotating holes was recently considered \cite{Rogatko-Szyplowska-2009}. The method outlined here will be of use to study the emission of massive fermions from rapidly-rotating higher-dimensional black holes.

This paper is organised as follows. In Sec.~\ref{sec-foundations} we introduce the basic equations. In Sec.~\ref{sec-a=0} we examine the non-rotating limit, and introduce a novel form for the spin-half spherical harmonics. In Sec.~\ref{sec-symmetries} we make explicit the symmetries of the eigenspectrum. In Sec.~\ref{sec-sig=mu} we derive exact expressions for the eigenvalues and eigenfunctions for the special cases $a\mu = \pm a\om$. In Sec.~\ref{sec-my-method} we define a spectral decomposition method for tackling the general case, which leads to a three-term recurrence relation. In Sec.~\ref{sec-asymptotics} we use the recurrence relation to investigate the asymptotic behaviour of the eigenvalue spectrum in the slow-rotation regime. In Sec.~\ref{sec-alt-methods} we describe some alternative methods used to find eigenvalue solutions, and highlight some errors in previous work. Numerical results are presented in Sec.~\ref{sec-results}. We examine the dependence of the eigenvalues on $a\mu$ and $a\om$, investigate the zeros of the eigenvalue spectrum, and present a gallery of eigenfunctions. We conclude with a brief discussion in Sec.~\ref{sec-conclusion}. Tables of eigenvalues, plus simple but accurate numerical fits to the data, are given in Appendix \ref{appendix-tables}.

\section{Basics\label{sec-foundations}}

In 1976 Chandrasekhar \cite{Chandrasekhar-1976} demonstrated how to separate variables for the massive Dirac field on the Kerr spacetime and shortly afterwards, Page \cite{Page-1976} extended the analysis to the Kerr-Newman spacetime. More recently, extensions to non-asymptotically flat \cite{Belgiorno-2008} and other related spacetimes \cite{Semiz-1992} have been considered. In such analyses, the $t$ and $\phi$ dependence may be factorised by using the ansatz $\Psi(t,r,\theta,\phi) = e^{-i \om t + i m \phi} \Psi_{\omega m} (r, \theta)$, and the Newman-Penrose method is applied~\cite{NP}. 
The four components of the wavefunction $\Psi_{\om m}$ can be expressed as products of two radial and two angular functions $\{ R_+(r), R_-(r), S_+(\theta), S_-(\theta) \}$ that obey coupled ODEs. 
The angular equations are:
%
\begin{eqnarray}
\frac{d \Sp}{d \theta} + \left( \frac{1}{2} \cot \theta - a \om \sin \theta + m \csc \theta \right) \Sp &= - \left( \lambda - a \mu \cos \theta \right) \Sm ,  \label{chandra1} \\
\frac{d \Sm}{d \theta} + \left( \frac{1}{2} \cot \theta + a \om \sin \theta - m \csc \theta \right) \Sm &= + \left( \lambda + a \mu \cos \theta \right) \Sp .  \label{chandra2}
\end{eqnarray}
Here, $a = J/M$ is the black hole angular momentum parameter, and $\om$ and $\mu$ are the frequency and mass of the state under consideration (N.B. the symbol $\sig \equiv -\om$ is also used for the frequency in the literature \cite{Suffern-1983}). 
$\Sp(\theta)$ and $\Sm(\theta)$ are known as mass-dependent spheroidal harmonics of spin one-half, and $\lambda$ is the eigenvalue. The eigenvalue can be regarded as the square root of a generalised squared total angular momentum~\cite{Batic-2005b}. The solutions depend on two continuous parameters, $a\om$ and $a\mu$. For a given $a\om$, $a\mu$, the eigenstates $\{ \Sp(\theta), \Sm(\theta), \lambda \}$ may be labelled by three discrete numbers: the angular momentum $j = 1/2, 3/2, \ldots$, the azimuthal component of the angular momentum $m = -j, -j+1, \ldots, j$ and the parity $\Par = \pm 1$. When this dependence is to be made explicit, we will write $S_{\pm} = _{s=\pm1/2}S_{jm\Par}^{(a\om, a\mu)}$ and $\lambda = \eig$. Equations (\ref{chandra1})--(\ref{chandra2}), and hence the eigenvalue spectrum and eigenfunctions, exhibit a number of symmetries (see Sec.~\ref{sec-symmetries}). Knowledge of the spectrum in the quadrant $a\omega >0$, $a \mu > 0$ is sufficient to determine the full spectrum.

The first-order equations (\ref{chandra1}) and (\ref{chandra2}) may be combined to obtain a second-order equation,
\begin{eqnarray}
\left[ \frac{d^2}{d \theta^2} + \cot \theta \frac{d}{d \theta} - \frac{(m-\frac{1}{2}\cos \theta)^2}{\sin^2 \theta} - \frac{1}{2} \right] S_- \quad \quad  &&  \nn \\
+\left[( a^2 \om^2 - a^2 \mu^2) \cos^2 \theta + a\om \cos \theta \right] S_- + a\mu \sin \theta S_+
&=& -[ \lambda^2 - a^2 \om^2 + 2a\om m] S_- ,   \label{chandrasekhar-page-eqn}
\end{eqnarray}
known as the \emph{Chandrasekhar-Page angular equation}. In the limit $\mu = 0$ this reduces to the well-known equation for the spin-weighted spheroidal harmonics (see for example Eq.~(2.1) in \cite{Berti-2005}, making the identifications $s=-1/2$ and ${}_sA_{lm} = \lambda^2 + 2a\omega m - a^2 \omega^2$). We find, in the case $a \mu \neq 0$, it is actually easier to analyse the coupled first-order equations directly.

There have been a number of studies of Eqs.~(\ref{chandra1}--\ref{chandrasekhar-page-eqn}) over the years \cite{Chandrasekhar-1983, Suffern-1983, Chakrabarti-1984, Kalnins-1992, Batic-2005a, Winklmeier-2008}. In Sec. \ref{sec-alt-methods} we discuss alternative approaches in some detail, to validate our method, and to identify some errors in the literature. 


\section{Exact Solutions for $a = 0$\label{sec-a=0}}
In the non-rotating limit, $a = 0$, the equations reduce to
\begin{eqnarray}
\frac{d \Sp}{d \theta} + \left( \frac{1}{2} \cot \theta + m \csc \theta \right) \Sp &= - \lambda \Sm \\
\frac{d \Sm}{d \theta} + \left( \frac{1}{2} \cot \theta - m \csc \theta \right) \Sm &= + \lambda \Sp .
\end{eqnarray}
Making the substitution 
\begin{equation}
\begin{bmatrix} \Sp \\ \Sm \end{bmatrix} = \begin{bmatrix} \coshalf & \sinhalf \\ - \sinhalf & \coshalf \end{bmatrix} \begin{bmatrix} \Tp \\ \Tm \end{bmatrix}
\label{half-angle}
\end{equation}
leads to coupled equations
\begin{equation}
\frac{d}{d\theta} \begin{bmatrix} \Tp \\ \Tm \end{bmatrix} + \begin{bmatrix} (m+1/2) \cot \theta & (m + \lambda + 1/2) \\ (m - 1/2 - \lambda) & - (m-1/2) \cot \theta \end{bmatrix} \begin{bmatrix} \Tp \\ \Tm \end{bmatrix}  = 0.
\end{equation}
These can be written in second-order form,
\begin{equation}
\frac{1}{\sin \theta} \frac{d}{d \theta} \left( \sin \theta \, \frac{d T_{\pm}}{d \theta} \right) - \frac{(m \pm 1/2)^2}{\sin^2 \theta} T_\pm + \lambda(\lambda + 1) T_\pm = 0,  
\end{equation}
which is simply the general Legendre equation. The eigenvalues are
\begin{equation}
\lam_{jm\Par}^{(0,0)} = \Par (j + 1/2)
\end{equation} 
where $\Par = \pm 1$ and $j$ is a half-integer.
The eigenfunctions are $T_\pm(\theta) \propto P_L^{m \pm 1/2} (\cos \theta)$, where $P_L^{m\pm1/2}$ is an associated Legendre polynomial obeying the Condon-Shortley phase convention, and $L = j + \Par/2$ is a non-negative integer. Let us define a set of solutions 
\begin{equation}
\begin{bmatrix} {}_{+1/2}Y_{jm\Par}(\theta) \\  {}_{-1/2}Y_{jm\Par}(\theta) \end{bmatrix} = 
\left( \frac{1}{2 \pi} \frac{(j - m)!}{(j+m)!} \right)^{1/2} \begin{bmatrix} \coshalf & \sinhalf \\ -\sinhalf & \coshalf \end{bmatrix} \begin{bmatrix} P_L^{m + 1/2} (x) \\ c_{jm\Par} P_L^{m - 1/2} (x)   \end{bmatrix}  \label{YSD-defn}
\end{equation}
where $x = \cos \theta$, $c_{jm\Par} = \Par(L + 1/2) - m$ and $L = j + \Par / 2$. 

It is straightforward to verify that these solutions are normalised so that
\begin{equation}
2 \pi \int_{\theta = 0}^{\pi}  d \theta  \sin \theta \, {}_{s} Y_{j m \Par}(\theta) \, {}_{s} Y_{j^\prime m \Par}(\theta) = \delta_{j j^\prime}
\end{equation}
and that they exhibit the symmetries 
\begin{eqnarray}
{}_sY_{jm\Par} (\theta) &=&  (-1)^{s - 1/2} \, {}_{s}Y_{jm,-\Par}(\theta)  ,  \label{sym1} \\
  &=& \Par (-1)^{m-1/2} \, {}_{-s}Y_{j,-m,-\Par} (\theta), \label{sym2} \\
  &=&  \Par (-1)^{j + m} \, {}_{-s}Y_{jm\Par}(\pi - \theta) .  \label{sym3} 
\end{eqnarray}
Note that by combining these symmetries, we may obtain a further four expressions of similar form. 

The solutions (\ref{YSD-defn}) are closely related to the \emph{spin-weighted spherical harmonics} of spin-weight half, ${}_{s=\pm1/2} \mathcal{Y}_{jm}^{(NP)}(\theta)$, first introduced by Newman and Penrose \cite{Newman-1966}. We make the identification
\begin{eqnarray}
{}_s\mathcal{Y}_{jm}^{(NP)}(\theta)   &=& (-1)^{m + s}  \, {}_{s} Y_{jm, \Par=+1} (\theta) \label{spin-weighted-new-form} \\
  &=& \left[ \frac{2j+1}{4\pi} \frac{(j+m)!}{(j+s)!} \frac{(j - m)!}{(j - s)!} \right]^{1/2} [\sin(\theta/2)]^{2j} \nn \\
  && \times \sum_n \begin{pmatrix} j - s \\ n \end{pmatrix} \begin{pmatrix} j + s \\ n + s - m \end{pmatrix} (-1)^{j-s-n} [\cot (\theta/2)]^{2n+s-m} \label{YNP-defn} 
\end{eqnarray}
The representation (\ref{YNP-defn}) was introduced by Goldberg \emph{et al.} (see \cite{Goldberg-1967}, eq. 3.1). The representation of ${}_s\mathcal{Y}_{jm}^{(NP)}$ in terms of Legendre polynomials (Eqs.~(\ref{YSD-defn}) and (\ref{spin-weighted-new-form})) does not seem to be well-known in the literature. These functions arise in the guise of spherical monogenics in geometric algebra \cite{Doran-1996, Doran-2003}, and we find that the representation (\ref{YSD-defn}) has many advantages. 

\section{Symmetries}\label{sec-symmetries}
The eigenvalue spectrum has the following symmetries
\begin{equation}
\lambda_{j,m,\Par}^{(a\omega, a\mu)} = - \lambda_{j,-m,-\Par}^{(-a\omega, a\mu)} =  - \lambda_{j,m,-\Par}^{(a\omega, -a\mu)} =  \lambda_{j,-m,\Par}^{(-a\omega, -a\mu)}  ,   \label{eig-sym}
\end{equation}
Hence knowledge of the spectrum in the quadrant $a\omega > 0$, $a \mu > 0$ is sufficient to determine the full spectrum.
In close correspondence with Eq.~(\ref{sym1})--(\ref{sym2}), the eigenfunctions have the following symmetries
\begin{eqnarray}
{}_sS_{jm\Par}^{(a\omega, a\mu)} (\theta) &=&  (-1)^{s - 1/2} \, {}_{s}S_{jm,-\Par}^{(a\omega, -a\mu)}(\theta)  ,  \label{S-sym1} \\
  &=& \Par (-1)^{m-1/2} \, {}_{-s}S_{j,-m,-\Par}^{(-a\omega, a\mu)} (\theta), \label{S-sym2} \\
  &=&  \Par (-1)^{j + m} \, {}_{-s}S_{jm\Par}^{(a\omega, a\mu)} (\pi - \theta) ,  \label{S-sym3} 
\end{eqnarray}
and combinations thereof. For example, combining (\ref{S-sym1}) and (\ref{S-sym2}) yields ${}_sS_{jm\Par}^{(a\omega, a\mu)} (\theta)  =  \Par (-1)^{s+m} \, {}_{-s}S_{j,-m,\Par}^{(-a\omega, -a\mu)} (\theta)$. 

\section{Exact solutions for $a\om = \pm a\mu$\label{sec-sig=mu}}
In the special case that $a\om = a \mu$ we find that transformation (\ref{half-angle}) yields the coupled equations
\begin{equation}
\frac{d}{d\theta} \begin{bmatrix} T_+ \\ T_- \end{bmatrix} + \begin{bmatrix}  (m+1/2) \cot \theta & (\lambda + m + 1/2 - a \om) \\ (m-1/2-\lambda - a\om) & -(m-1/2) \cot \theta \end{bmatrix} \begin{bmatrix} T_+ \\ T_- \end{bmatrix}
 = 0
\end{equation}
Again, the solutions are associated Legendre polynomials, $T_\pm \propto P_L^{m\pm1/2}$, where $L$ is a non-negative integer. The normalised solutions are
\begin{equation}
\begin{bmatrix} {}_{+1/2}S_{jm\Par}^{(a\om, a\om)} (\theta) \\  {}_{-1/2}S_{jm\Par}^{(a\om, a\om)} (\theta) \end{bmatrix} = 
A  \begin{bmatrix} \coshalf & \sinhalf \\ -\sinhalf & \coshalf \end{bmatrix} \begin{bmatrix} P_L^{m + 1/2} (x) \\ b_{jm\Par} P_L^{m - 1/2} (x)   \end{bmatrix}  \label{Smu-neg-defn}
\end{equation}
where $L = j + \Par / 2$ and $b_{jm\Par} = [(L+1/2)^2 - m^2] / (\lambda + m + 1/2 - a\om)$ and the normalisation factor is
\begin{equation}
A^2 = \frac{2L+1}{4 \pi} \frac{(L-m-1/2)!}{(L+m+1/2)!} \left( 1 + \frac{\Par(m-a\om)}{\sqrt{ (L+1/2)^2 - 2ma\om + a^2 \om^2 }} \right) .
\end{equation}
The corresponding eigenvalue is
\begin{equation}
\lam_{jm\Par}^{(a\om, a\om)} = - \frac{1}{2} + \Par \sqrt{(L+1/2)^2 - 2ma\om + a^2 \om^2}.  
\label{lam-exact}
\end{equation}
In the special case $m = -j, P = -1$ we find $b_{|m|,m,-1} = 1$ and $\lam_{|m|,m,-1} = -1/2 - j - a\om$.

The solution for $a \om = -a \mu$ can be found by considering the symmetry of equations (\ref{chandra1}) and (\ref{chandra2}) under the simultaneous transformations
\begin{equation}
a\mu \rightarrow - a\mu, \quad \lambda \rightarrow -\lambda, \quad {}_sY_{jm\Par}(\theta) \rightarrow (-1)^{s-1/2} {}_sY_{jm -\Par}(\theta)
\end{equation}
Hence the solutions are
\begin{equation}
\begin{bmatrix} {}_{+1/2}S_{jm\Par}^{(a\om, -a\om)} (\theta) \\  {}_{-1/2}S_{jm\Par}^{(a\om, -a\om)} (\theta) \end{bmatrix} = 
B \begin{bmatrix} \coshalf & -\sinhalf \\ \sinhalf & \coshalf \end{bmatrix} \begin{bmatrix} P_L^{m + 1/2} (x) \\ b_{jm\Par}^{(a\om, -a\om)} P_L^{m - 1/2} (x)   \end{bmatrix}  \label{Smu-pos-defn}
\end{equation}
where now $L = j - \Par / 2$ and $b_{jm\Par}^{(a\om, -a\om)} = - [(L+1/2)^2 - m^2] / (-\lam_{jm\Par}^{(a\om,-a\om)} + m + 1/2 - a\om)$ and the normalisation factor is 
\begin{equation}
B^2 = \frac{2L+1}{4 \pi} \frac{(L-m-1/2)!}{(L+m+1/2)!} \left( 1 - \frac{\Par(m-a\om)}{\sqrt{ (L+1/2)^2 - 2ma\om + a^2 \om^2 }} \right) .
\end{equation}
The corresponding eigenvalue is
\begin{equation}
\lam_{jm\Par}^{(a\om, -a\om)} = \frac{1}{2} + \Par \sqrt{(L+1/2)^2 - 2 m a \om + a^2 \om^2}.  \label{lam-exact-2}
\end{equation}

\section{A Spectral Decomposition Method\label{sec-my-method}}
An obvious next step is to seek an expansion for the spheroidal harmonics in the basis of spherical harmonics (\ref{YSD-defn}). This was the approach followed by Chakrabarti \cite{Chakrabarti-1984}, who applied the method to the second-order Chandrasekhar-Page equation (\ref{chandrasekhar-page-eqn}) to find a five-term recurrence relation. Here we apply the method directly to the first-order equations themselves, to recover a three-term recurrence relation. The key advantage of a three-term relation is that it can be solved using robust continued-fraction methods.

Let us begin with the ansatz
\begin{equation}
{}_{s}S_{jm\Par}^{(a\om, a\mu)} (\theta) = \sum_{k^\prime=|m|}^\infty {}_sc_{k^\prime m\Par (j)}^{(a\om, a\mu)} \, {}_{s}Y_{k^\prime m\Par}(\theta)   \label{expansion-ansatz}
\end{equation}
where ${}_sc_{k^\prime m\Par (j)}^{(a\om, a\mu)}$ are expansion coefficients to be determined, and ${}_{s}Y_{k^\prime m\Par}(\theta)$ were defined in (\ref{YSD-defn}). The symmetries (\ref{sym3}) and (\ref{S-sym3}) imply that the coefficients are related by
\begin{equation}
{}_sc_{k^\prime m\Par(j)}^{(a\om, a\mu)} = (-1)^{j - k^\prime} {}_{-s}c_{k^\prime m\Par (j)}^{(a\om, a\mu)} \label{csym}
\end{equation}
For clarity, we neglect the least-relevant indices, writing ${}_{\pm 1/2}c_{k^\prime m\Par (j)}^{(a \sig, a\mu)} \equiv c_{k^\prime (j)}^{(\pm)}$ and $ {}_{\pm 1/2}  Y_{k^\prime m\Par} \equiv Y_{k^\prime}^{(\pm)}$. Substituting ansatz (\ref{expansion-ansatz}) into equation (\ref{chandra1}) and using (\ref{csym}) leads to
\begin{equation}
\sum_{k^\prime = |m|}^\infty \left( - \Par (k^\prime + 1/2) Y_{k^\prime}^{(-)} - a \om \sin \theta Y_{k^\prime}^{(+)} + (-1)^{j - k^\prime} ( \lam - a \mu \cos \theta ) Y_{k^\prime}^{(-)} \right)  c^{(+)}_{k^\prime (j) } = 0 . 
\end{equation}
If we now multiply by $Y_{k}^{(-)}$ and integrate we obtain the matrix eigenvalue equation
\begin{equation}
\sum_{k^\prime=|m|}^\infty A_{k k^\prime} b_{k^\prime}  =  \lambda b_k   \label{eigval-eqn}
\end{equation}
where 
\begin{eqnarray}
A_{k k^\prime} &=& (-1)^{j - k^\prime} (k+1/2) \Par \delta_{k k^\prime} + (-1)^{j - k^\prime} a\om \, D_{kk^\prime}^{(1)} + a\mu \, C_{kk^\prime}^{(1)} ,  \label{rec-relation-1} \\
b_k &=& (-1)^{j - k} c^{(+)}_{k(j)} = c^{(-)}_{k(j)},
\end{eqnarray}
and
\begin{eqnarray}
C_{kk^\prime}^{(1)} &=& 2 \pi \int_0^\pi d\theta \sin \theta \, Y_k^{(-)}(\theta) \, \cos \theta \, Y_{k^\prime}^{(-)}(\theta) , \\
 &=& \left( \frac{2 k^\prime + 1}{2 k + 1} \right)^{1/2} \left< k^\prime \, 1 \, m \, 0 \, | \, k \, m \right> \left< k^\prime \, 1 \, \tfrac{1}{2} \, 0 \, | \, k \, \tfrac{1}{2} \right>,   \label{bigC-defn} \\
D_{kk^\prime}^{(1)} &=& 2 \pi \int_0^\pi d\theta \sin \theta \, Y_k^{(-)}(\theta) \, \sin \theta \, Y_{k^\prime}^{(+)} (\theta) , \\
 &=&  \Par \left( 2 \, \frac{2 k^\prime + 1}{2 k + 1} \right)^{1/2} \left< k^\prime \, 1 \, m \, 0 \, | \, k \, m  \right> \left< k^\prime \, 1 \, -\tfrac{1}{2} \, 1 \, | \, k \, \tfrac{1}{2} \right> .  \label{bigD-defn}
\end{eqnarray}
The Clebsch-Gordan coefficients are only non-zero for $k^\prime - 1 \le k \le k^\prime+1$. Hence $A_{kk^\prime}$ is a tridiagonal matrix, and the eigenvalues $\lambda$ and vectors $b_k$ can be found via linear algebra routines. Alternatively, we can obtain a three-term recurrence relation for the expansion coefficients $c_{k(j)}^{(+)}$. Using expressions for the Clebsch-Gordan coefficients given in Appendix \ref{appendix-clebsch} it is straightforward to show that
\begin{eqnarray}
\alpha_k b_{k+1} + \beta_k b_k + \gamma_k b_{k-1} = 0,   \quad \quad \quad \quad  k = |m|, |m| + 1, \ldots
\label{recurr}
\end{eqnarray}
where
\begin{eqnarray}
\alpha_k & = &  \left( a\mu + \eps_k a \om \right) \frac{\sqrt{(k+1)^2 - m^2}}{2 (k+1)} , \\
\beta_k  & = & \eps_k (k + 1/2) \left( 1 - \frac{ a \om m }{k(k+1)}  \right) + \frac{a \mu m}{2 k (k+1)} - \lambda  , \\
\gamma_k & = & \left( a \mu - \eps_k a \om \right)  \frac{\sqrt{k^2 - m^2}}{ 2 k }  .
\label{gamk}
\end{eqnarray}
and $\eps_k = (-1)^{j - k} \Par$. Note that $j,k$ and $m$ are half-integers. The advantage of a three-term recurrence relation is that it can be solved via continued-fraction methods. 
To obtain numerical results, we followed the approach outlined in \cite{Leaver-1985}, and used a rescaling algorithm described in \cite{NumericalRecipes}. The basic idea is to separate out the eigenvalue by writing $\beta_k = \tilde{\beta}_k - \lambda$, and write down an expression for the ratio of consecutive terms
\begin{equation}
\frac{b_{k}}{b_{k-1}} = -\frac{\gamma_k}{\tilde{\beta}_k - \lambda + \alpha_k \left(\frac{b_{k+1}}{b_k}\right)}.
\end{equation}
The series should converge and this condition gives the possible values for the eigenvalues. In practice, we write the eigenvalue as a continued fraction
\begin{equation}
\lambda = \beta_1 - \frac{\alpha_1\gamma_3}{\beta_3 - \lambda - \frac{\alpha_3\gamma_5}{\beta_5 - \lambda - \frac{\alpha_5\gamma_7}{\beta_7 - \lambda-\cdots}}}
\label{cntfrac}
\end{equation}
This fraction can be continued to a certain level, and then we use the fact that the ratio $b_{k+1}/b_k \rightarrow 0$ as $k\rightarrow \infty$ to ignore subsequent terms. This is an equation of the form $\lambda = g(\lambda)$ which can be solved iteratively, by starting with a guess for $\lambda$, evaluating the right had side of~(\ref{cntfrac}), using this as a new estimate of $\lambda$ and repeating.

\section{Asymptotics\label{sec-asymptotics}}
Below we show that the new three-term recurrence relation (\ref{recurr}) may be used to study the eigenspectrum in the small-$a\omega, a\mu$ limit. To understand the opposite limit, $a \omega, a\mu \rightarrow \infty$, we recall some key results in the literature.

\subsection{Small $a\omega$ and $a\mu$}
For small values of $a \omega$ and $a \mu$ one may express the separation constant $\lambda$ as a power series. We start with the continued fraction equation in the form  
\begin{equation}
\beta_j - \frac{\alp_{j-1} \gam_j}{\beta_{j-1} - \ldots} \ldots \left( \frac{\alp_{|m|} \gam_{|m|+1}}{\beta_{|m|}} \right) = \frac{\alp_j \gam_{j+1}}{\beta_{j+1} - \ldots} \frac{\alp_{j+1}\gam_{j+2}}{\beta_{j+2} - \ldots} \ldots
\end{equation}
and expand the separation constant as a Taylor series,
\begin{equation}
\lam_{jm\Par} =  \sum_{p=0}^\infty \sum_{q=0}^\infty \Lambda^{(jm\Par)}_{pq} (a\omega)^{p} (a\mu)^{q}
\label{eq:small-asymptotics}
\end{equation}
By grouping together like powers of $a\omega$ and $a \mu$ we obtain expansion coefficients,
\begin{eqnarray}
\Lambda_{00} &=& \phantom{-\tfrac12} \Par(j+1/2)  \label{Lam-coeff-begin} \\
\Lambda_{10} &=& -\tfrac12 \Par m K_0^+(j) \\
\Lambda_{01} &=& \phantom{-}\tfrac12 m K_0^-(j) \\
\Lambda_{20} &=& \phantom{-\tfrac12} \Par \left[ H(j) + H(j+1) \right] \\
\Lambda_{02} &=& \phantom{-\tfrac12} \Lambda_{20}  \\
\Lambda_{30} &=& \phantom{-}\tfrac12 \Par m \left[ K_1^+(j) H(j) + K_1^+(j+1)H(j+1) \right]  \\
\Lambda_{03} &=& \phantom{-}\tfrac12  m  \left[ K_1^-(j) H(j) - K_1^-(j+1) H(j+1) \right] \\
\Lambda_{11} &=& \phantom{-} 2 \left[ H(j+1) - H(j) \right] \\
\Lambda_{21} &=& \phantom{-}\tfrac12 m \left[ \left( 2K^+_1(j+1) - K^-_1(j+1) \right) H(j+1) - \left( 2K^{+}_1(j) - K^{-}_1(j) \right) H(j) \right] \\
\Lambda_{12} &=& \phantom{-}\tfrac12 m \Par \left[ \left( K_1^+(j+1) - 2K_1^-(j+1) \right) H(j+1) + \left(K_1^+(j) - 2K_1^-(j) \right) H(j)  \right]   \label{Lam-coeff-end}
\end{eqnarray}
where (for compactness) we have defined the functions
\begin{eqnarray}
H(k) &=& (k^2 - m^2) / (8 k^3) \\
K_0^\pm(k) &=& 1/k  \pm  1/(k+1) \\
K_1^\pm(k) &=& 1/[(k+1)(k-1)] \pm 1 / k^2 .
\end{eqnarray}

If desired, the series may be continued to higher orders with the aid of a symbolic algebra package. It is straightforward to confirm that the expansion coefficients are consistent with the exact eigenvalues for $a \omega = \pm a \mu$ presented in Sec. \ref{sec-sig=mu} and expansions for the massless case ($a\mu = 0$) given in, e.g., \cite{Berti-2005}. In Sec.~\ref{sec-alt-methods} we validate against an alternative series expansion in powers of $(a \omega \pm a\mu)$ obtained in \cite{Suffern-1983, Batic-2005a}.

\subsection{Large $a \omega$, massless $a\mu = 0$ }
In the massless case ($a \mu = 0$), asymptotic results for the eigenvalue in the large-$a\omega$ limit were obtained by Breuer, Ryan and Waller \cite{Breuer}. In our notation,
\begin{equation}
 \lambda^2 = 2 (q - m) a\omega +  \mathcal{A}_0 + \mathcal{A}_1 / (a\omega) + \mathcal{A}_2 / (a\omega)^2 + \mathcal{A}_3 / (a\omega)^3  + \mathcal{O}((a\omega)^{-4})
 \label{eq:large-asymptotics}
\end{equation}
where 
\begin{eqnarray}
\mathcal{A}_0 &=&  -  \tfrac{1}{2} \left[ q^2 - m^2 + 2s + 1  \right] \\
\mathcal{A}_1 &=&  - \tfrac{1}{8} \left[ q^3 - m^2 q + q - s^2(q+m) \right] \\
\mathcal{A}_2 &=&  -\frac{1}{64} \left( 5q^4 - (6m^2-10) q^2 + m^4 - 2m^2 - 4s^2(q^2-m^2-1) + 1 \right)  
\end{eqnarray}
and $s = -1/2$ and $\mathcal{A}_3$ is given in \cite{Breuer}. 
Breuer \emph{et al.} leave the parameter $q$ undetermined. Casals and Ottewill \cite{CasalsOttewill} showed that $q$ may be obtained by counting the number of zeros of the solution. We obtain the best match between (\ref{eq:large-asymptotics}) and our numerical results when we take $q = j + z$, where $z = 0$ if $|j-m|$ is even and $z=1$ otherwise (see Fig.~\ref{fig-eigenvalues-mu0}). However, we are unable to find a suitable asymptote for the $m = j$ mode.

\subsection{Large $a \omega$ and $a\mu$ }
The solutions with $a\mu=0$ are a special case of the asymptotic behaviour. This is most easily seen from the Chandrasekhar-Page equation, Eq.~(\ref{chandrasekhar-page-eqn}). When $a\mu=0$, the terms quadratic in $a\omega$ can be combined into a single term $-a^2\omega^2\sin^2\theta S_-$. As a result, if the eigensolutions are confined to a region near $\theta=0$ of size $\Delta\theta \lesssim 1/\sqrt{a\omega}$, this term is actually $O(a\omega)$ and hence we expect $\lambda^2 \sim a\omega$. It is clear from the results of Breuer \emph{et al.}~\cite{Breuer}, given in Eq.~(\ref{eq:large-asymptotics}), that this is indeed the case for the asymptotic $a\mu=0$ solutions.

\begin{figure}
\begin{tabular}{cc}
\includegraphics[width=0.5\textwidth]{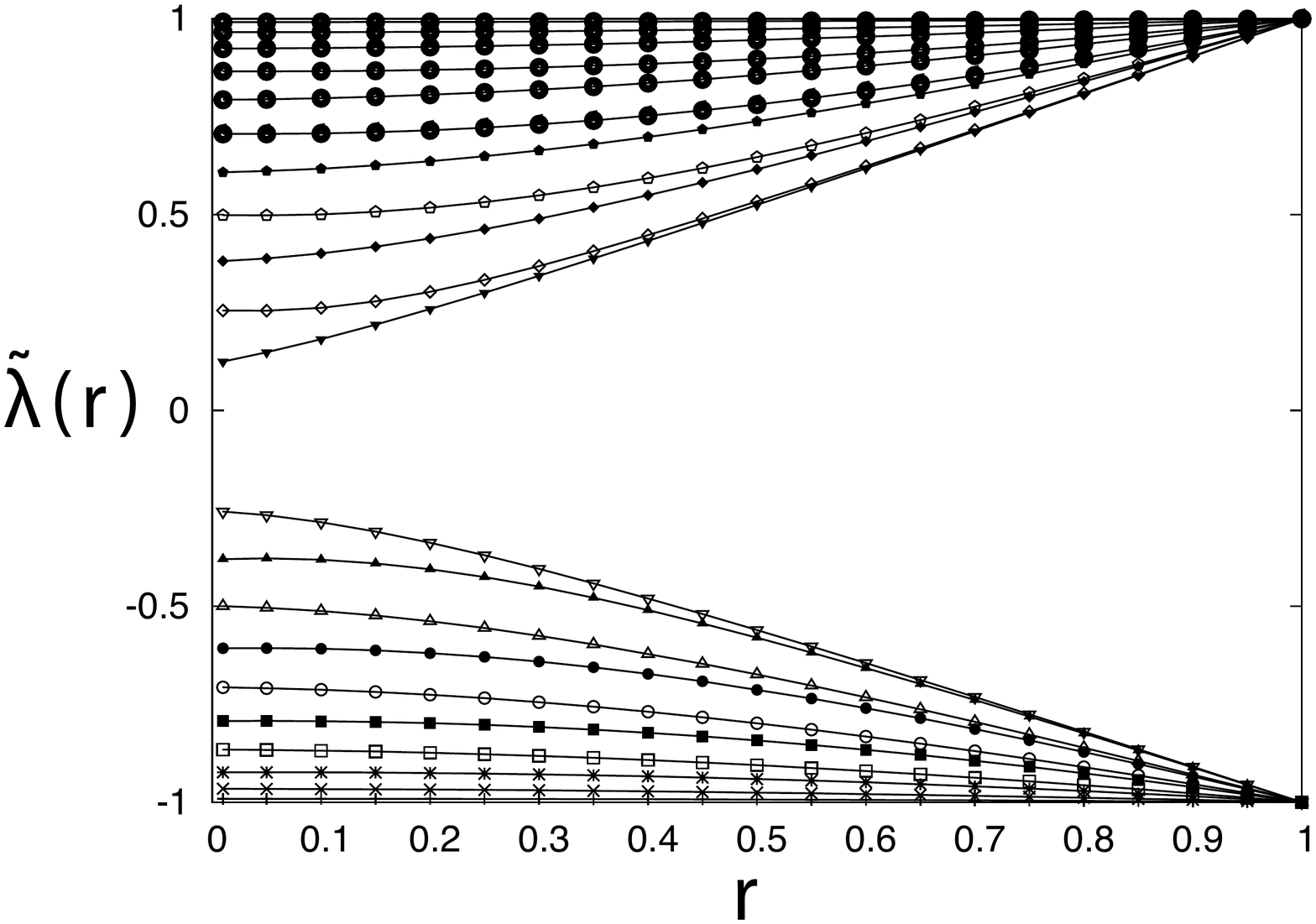}&
\includegraphics[width=0.5\textwidth]{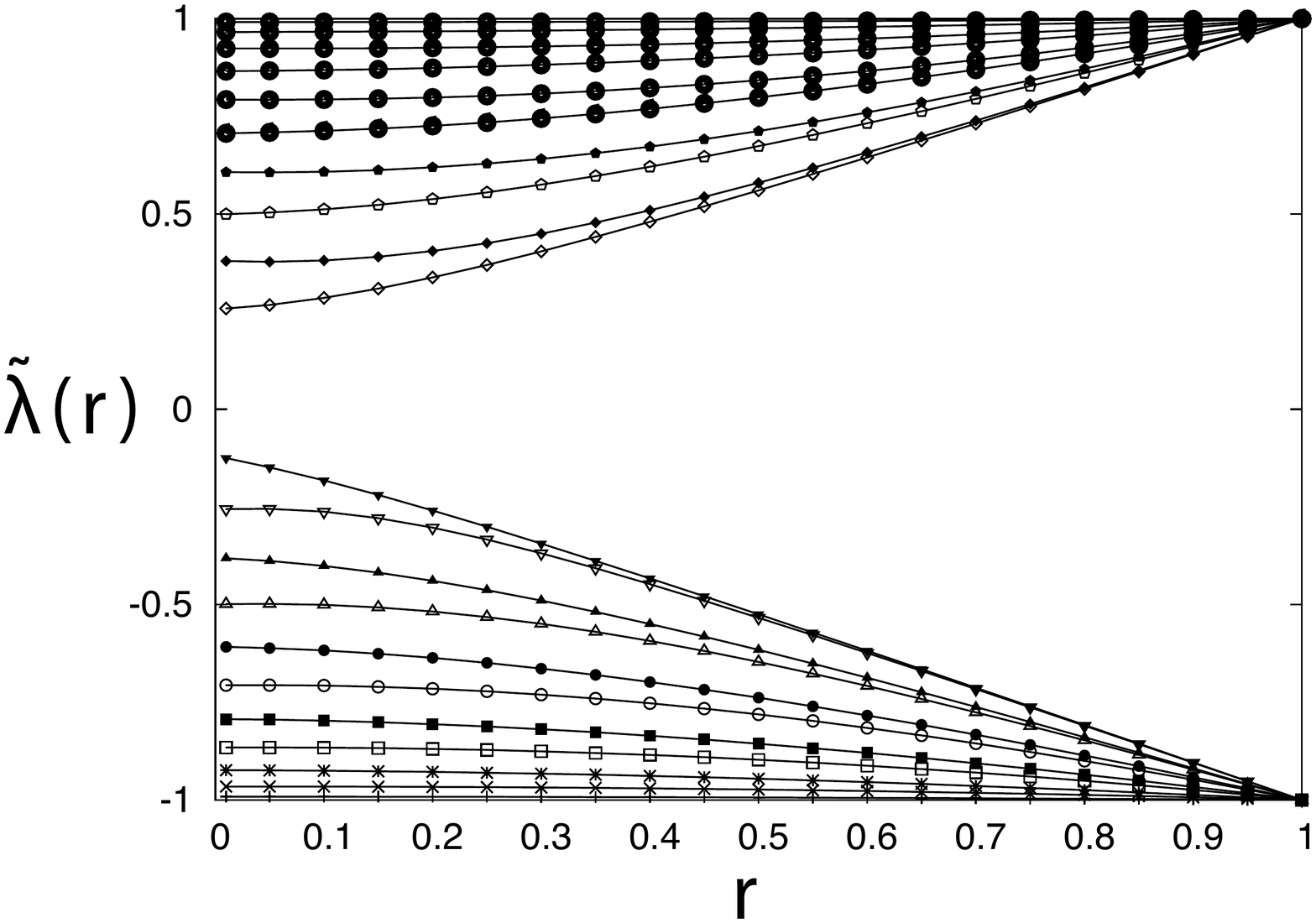}
\end{tabular}
\caption[]{\emph{Solutions to the asymptotic problem}. Lowest modes of the solution to the asymptotic recurrence relation, Eqs~(\ref{ASYMrecurr})--(\ref{ASYMgamk}), for $m=1/2$ (left panel) and $m=-1/2$ (right panel), as a function of the fixed mass-to-frequency ratio $r=\mu/\omega$. Modes with $\tilde{\lambda}(r) > 0$ have parity $\Par=1$ and those with $\tilde{\lambda}(r) < 0$ have parity $\Par=-1$.}
\label{fig:asymp}
\end{figure}

If instead we consider the limit $a\omega \rightarrow \infty$, with the ratio $r=\mu/\omega\neq0$ fixed, there is an additional quadratic term, $-a^2\omega^2r^2\cos^2\theta S_-$, in Eq.~(\ref{chandrasekhar-page-eqn}). It is not possible for both this term and the $-a^2\omega^2\sin^2\theta S_-$ term to be small simultaneously, and so we expect that $\lambda^2 \sim a^2\omega^2$ in this case. This is borne out by our numerical results. Writing $\lambda = a\omega \tilde{\lambda}(r) + \cdots$, the asymptotic solutions $\tilde{\lambda}(r)$ may be found by taking the limit $a\omega \rightarrow \infty$ of the recurrence relation, Eqs~(\ref{recurr})--(\ref{gamk})
\begin{eqnarray}
\tilde{\alpha}_k b_{k+1} + \tilde{\beta}_k b_k + \tilde{\gamma}_k b_{k-1} = 0,   \quad \quad \quad \quad  k = |m|, |m| + 1, \ldots
\label{ASYMrecurr}
\end{eqnarray}
where
\begin{eqnarray}
\tilde{\alpha}_k & = &  \left( r + \eps_k \right) \frac{\sqrt{(k+1)^2 - m^2}}{2 (k+1)} , \\
\tilde{\beta}_k  & = & -\frac{\eps_k (k + 1/2)m}{k(k+1)} + \frac{m}{2 k (k+1)} - \tilde{\lambda}(r)  , \\
\tilde{\gamma}_k & = & \left( r - \eps_k \right)  \frac{\sqrt{k^2 - m^2}}{ 2 k }  .
\label{ASYMgamk}
\end{eqnarray}
and $\eps_k = (-1)^{j - k} \Par$ as before. We can solve this recurrence using the same method outlined in Section~\ref{sec-my-method}. The results are shown in Fig.~\ref{fig:asymp} for $0 < r \lesssim 1$. We note that as $r \rightarrow 1$, all the positive parity modes converge at $\tilde{\lambda}=1$ and the negative parity modes converge at $\tilde{\lambda}=-1$, which is consistent with the $a\omega \rightarrow \infty$ limit of the exact solution for $r=1$, Eq.~(\ref{lam-exact}). We have verified numerically that the eigenvalues of the full problem do indeed have a linear slope asymptotically, and that this slope is correctly predicted by the asymptotic solutions shown in Fig.~\ref{fig:asymp}.

\section{Alternative Methods\label{sec-alt-methods}}
As discussed briefly in Sec.~\ref{sec-foundations}, alternative methods for finding eigensolutions exist in the literature, although some mistakes are also present. Here we seek to review and clarify the present situation.

Suffern, Fackerell and Cosgrove (hereafter SFC) \cite{Suffern-1983} showed that the eigenfunctions could be expressed as an infinite series of hypergeometric ${}_2F_1$ functions, and derived a three-term recurrence relation for the series coefficients (\cite{Suffern-1983}, Eq. (16)). They showed that the series terminates for the special case $a \omega = - a\mu$, and derived an exact result compatible with (\ref{lam-exact-2}) in this case. Their three-term relation can be solved via the continued-fraction method to determine $\lambda$. We have verified that the eigenvalues calculated via SFC's recurrence relation agree with the eigenvalues calculated via (\ref{eigval-eqn}).  A key result presented in SFC was a series expansion of the form 
\begin{equation}
\lambda = \sum \sum C_{rs} (a\sig - a\mu)^r (a \sig + a\mu)^s .   \label{SFC-expansion}
\end{equation}
where $\sig = -\omega$. Unfortunately, the expansion coefficients presented in Table I of \cite{Suffern-1983} are not all correct (for example, diagonal elements $C_{nn}$ should be zero \cite{Batic-2005a}). In the second and fourth tables of Table II of \cite{Suffern-1983}, the eigenvalues are only accurate to one decimal place for $m = \pm 1/2$ and $j=3/2$ and $j=1/2$.

A five-term recurrence relation was calculated by Chakrabarti \cite{Chakrabarti-1984}; however, we believe the result should be treated with caution because the numerical eigenvalues presented are incorrect. For example, in Tables I, IIa, IIb and III in \cite{Chakrabarti-1984}, the eigenvalues are wrong, except for the special case $a\mu = 0$. There seems to be a tacit assumption made that the spectrum is symmetric about zero (i.e.~that if $\lambda$ is an eigenvalue then so is $- \lambda$), which is not the case for $a\mu \neq 0$. The correct symmetries of the spectrum are given in Eq.~(\ref{eig-sym}). 

Kalnins and Miller \cite{Kalnins-1992} also obtained a three-term recurrence relation, and a series expansion in powers of $a$, viz. $\lambda = \sum_{r=0} \lambda_r a^r$. 
Ranganathan \cite{Ranganathan-2006} presented exact solutions for the special case $\omega = \pm \mu$, $m = \pm j$, which are compatible with those given in Sec.~\ref{sec-sig=mu} and in SFC \cite{Suffern-1983}. 

Batic, Schmid and Winklmeier \cite{Batic-2005a} (hereafter BSW) recently showed that the eigenvalues of (\ref{chandrasekhar-page-eqn}) satisfy a first order partial differential equation,
\begin{equation}
(\tilde{\mu} - 2 \tilde{\nu} \lam ) \frac{\partial \lam}{\partial \tilde{\mu}} + (\tilde{\nu} - 2 \tilde{\mu} \lam) \frac{\partial \lam}{\partial \tilde{\nu}} + 2 m \tilde{\mu} + 2 \tilde{\mu} \tilde{\nu} = 0,  \label{pde-bsw}
\end{equation}
where $\tilde{\mu} = a\mu$ and $\tilde{\nu} = -a\omega$. 
Using this PDE, BSW found a method for calculating the series coefficients $C_{rs}$ in SFC's expansion (\ref{SFC-expansion}). BSW point out that the method applied by SFC to determine the series coefficients is plagued by divide-by-zero problems at certain orders ($r,s$). BSW describe a method for avoiding the problem, 
allowing the series expansion to be taken to arbitrary order. We have checked that the expansion coefficients given in \cite{Batic-2005a} are in full agreement with our expansion, Eq. (\ref{Lam-coeff-begin})--(\ref{Lam-coeff-end}).  The non-linear PDE (\ref{pde-bsw}) may be solved by the method of characteristics, and BSW show that the characteristic equations can be reduced to a Painlev\'e III equation \cite{Batic-2005a}.

Recently, Batic and Nowakowski~\cite{Batic-Nowakowski-2008} (hereafter BN) derived an ordinary differential equation for the eigenvalues at fixed mass-to-frequency ratio $r = \mu / \om$ , using analytic perturbation theory. This equation was
\begin{equation}
\frac{d \lambda}{d a } =  \frac{ 2 ( a \om - m_j ) ( 2 \lambda a \om - a \mu )}{a (4 \lambda^2 - 1)}.  \label{BN-eq}
\end{equation}
Unfortunately, this equation is not correct (see discussion below). 
In the corrected calculation, the eigenvalue is given once again by a partial differential equation, so a solution is not so easily obtained.

We have found that it is possible to adapt the approach of BN in order to derive an alternative recurrence relation for the angular eigenvalues. Following the same notation as BN, we write $S_+ =\tilde{g}_1/\sqrt{\sin\theta}$, $S_- =\tilde{g}_2/\sqrt{\sin\theta}$, $k=-m$, $m_e = \mu$ and $M=a$. In deriving the ODE (\ref{BN-eq}), BN introduced functions $U = \tilde{g}_1^2 + \tilde{g}_2^2$, $V = \tilde{g}_2^2 - \tilde{g}_1^2$ and $W = 2\tilde{g}_1\,\tilde{g}_2$, which obey the differential equations
\begin{eqnarray}
U'(\theta) &=& -2f(\theta) V(\theta) + 2 M m_e \cos\theta W(\theta), \qquad f(\theta) = M\om\sin\theta + \frac{k}{\sin\theta} , \label{BN35}\\
V'(\theta) &=& -2f(\theta) U(\theta) + 2\lambda W(\theta) \label{BN36}\\
W'(\theta) &=& 2M m_e \cos\theta U(\theta) - 2\lambda V(\theta)\label{BN37}
\end{eqnarray}
These are equations (3.5)--(3.7) of BN, but we have corrected a factor of 2 in equation~(\ref{BN35}). BN state correctly that ${\rm d}\lambda/{\rm d}M = m_e I_1 + \omega I_2$, where
\begin{equation}
I_1 = \int_0^\pi \cos\theta V(\theta)\,{\rm d}\theta, \qquad I_2 = \int_0^\pi \sin\theta W(\theta)\,{\rm d}\theta.
\end{equation}
BN went on to derive three equations for $I_1$, $I_2$ and a third integral, $I_3 = \int_0^\pi \cos^2\theta U(\theta) \,{\rm d}\theta$. However, the sign of $\omega$ on the left hand side of their third equation (number 3.15) was wrong, with the consequence that they appeared to obtain three independent equations, whereas in reality the third equation was a linear combination of the first two. The fact that their original solution was not correct was indicated by the consequence that $I_3 \equiv 0$, while $I_3$ is the integral of the square of a real function and must therefore be greater than zero.

It is possible to adapt and extend their approach as follows. We introduce two families of integrals, which generalise the integrals used in BN
\begin{equation}
J_n = 2 \int_0^{\pi} \cos^n\theta\,\,\tilde{g}_2^2(\theta)\,\,{\rm d}\theta, \qquad L_n = 2 \int_0^{\pi} \sin\theta\,\cos^{2n}\theta\,\,\tilde{g}_1(\theta)\tilde{g}_2(\theta)\,\,{\rm d}\theta,
\end{equation}
Due to the symmetry of the eigenfunctions $\tilde{g}_1(\pi - \theta) = \tilde{g}_2(\theta)$, and the fact that they are normalised such that $\int_0^{\pi} \tilde{g}_i^2 \,{\rm d}\theta = 1$, we see that $J_0=2$, $J_1=I_1$, $L_1=I_2$ and $J_2=I_3$. Multiplying Eq.~(\ref{BN35}) by $\sin\theta \,\cos^{2n-1}\theta$ and integrating we find
\begin{equation}
(2n-1)J_{2n-2}-2nJ_{2n} = 2Mm_eL_n + 2M\om J_{2n+1} - 2(M\omega + k)J_{2n-1} .
\label{rec1}
\end{equation}
Multiplying Eq.~(\ref{BN36}) by $\sin\theta \,\cos^{2n}\theta$ and integrating gives
\begin{equation}
2nJ_{2n-1}-(2n+1)J_{2n+1}=2M\om J_{2n+2}-2(M\om+k) J_{2n} + 2\lambda L_n.
\end{equation}
Finally, multiplying Eq.~(\ref{BN37}) by $\cos^{2n+1}\theta$ and integrating gives
\begin{equation}
(2n+1) L_n = 2Mm_eJ_{2n+2}-2\lambda J_{2n+1}.
\end{equation}
These three relations together can be solved recursively to obtain all of the integrals. For eigenvalues of the system, $J_n, L_n \rightarrow 0$ as $n\rightarrow \infty$, and this can be used to solve the recurrences. Combining the three equations gives a five-term recurrence for $\lambda$
\begin{eqnarray}
(4n^2-1)J_{2n-2}+2(2n+1)(M\om+k)J_{2n-1}-2n(2n+1)J_{2n} \nonumber \\ +(2(2n+1)\lambda-2M\om)J_{2n+1} -4M^2 m_e^2 J_{2n+2} = 0 \label{BNrec1}
\end{eqnarray}
or a four-term recurrence that now depends on $\lambda^2$
\begin{eqnarray}
2n(2n+1)J_{2n-1}+2(2n+1)(M\om+k)J_{2n} - ((2n+1)^2-4\lambda^2)J_{2n+1} \nonumber \\
 - (2(2n+1)M\om + 4Mm_e\lambda)J_{2n+2}=0 . \label{BNrec2}
\end{eqnarray}
It is clear from the above that it is possible to use the approach of BN to derive the eigenvalues, without resorting to a spectral decomposition. However, the result is a recurrence relation and the nice analytic solution for the eigenvalue derived in the original version of their paper is lost. While it may be possible to modify their approach further to obtain the eigenvalue exactly, we have so far been unable to do this. The recurrences in Eqs.~(\ref{BNrec1})--(\ref{BNrec2}) are more difficult to work with than the three-term recurrence in Eq.~(\ref{recurr}) and so we have used expression~(\ref{recurr}) to compute all the results presented in Section~\ref{sec-results}.

\section{Results\label{sec-results}}
In this section we present some numerical results for the eigenvalues and eigenfunctions of equations (\ref{chandra1}) and (\ref{chandra2}). These were computed using the three-term recurrence relation, Eqs.~(\ref{recurr})--(\ref{gamk}), and the series expansion in spin-half spherical harmonics (\ref{expansion-ansatz}). We have cross-checked against results from the other techniques (Sec.~\ref{sec-alt-methods}). As before we separate variables using the ansatz $\Psi(t,r,\theta,\phi) = e^{-i\omega t} e^{i m \phi} \Psi_{\omega m}(r, \theta)$.

\subsection{Eigenvalues}
\begin{figure}
\begin{center}
\includegraphics[width=12cm]{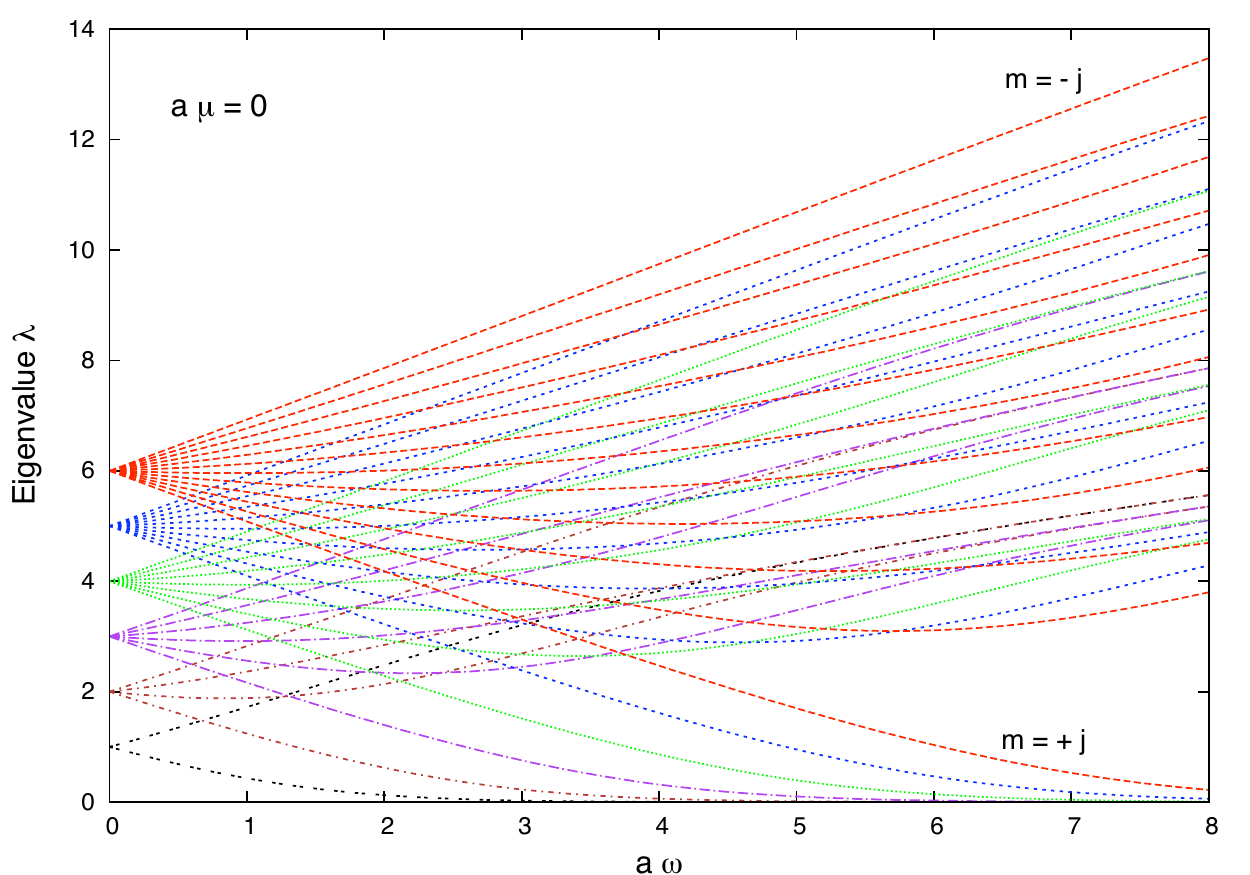}
\includegraphics[width=8cm]{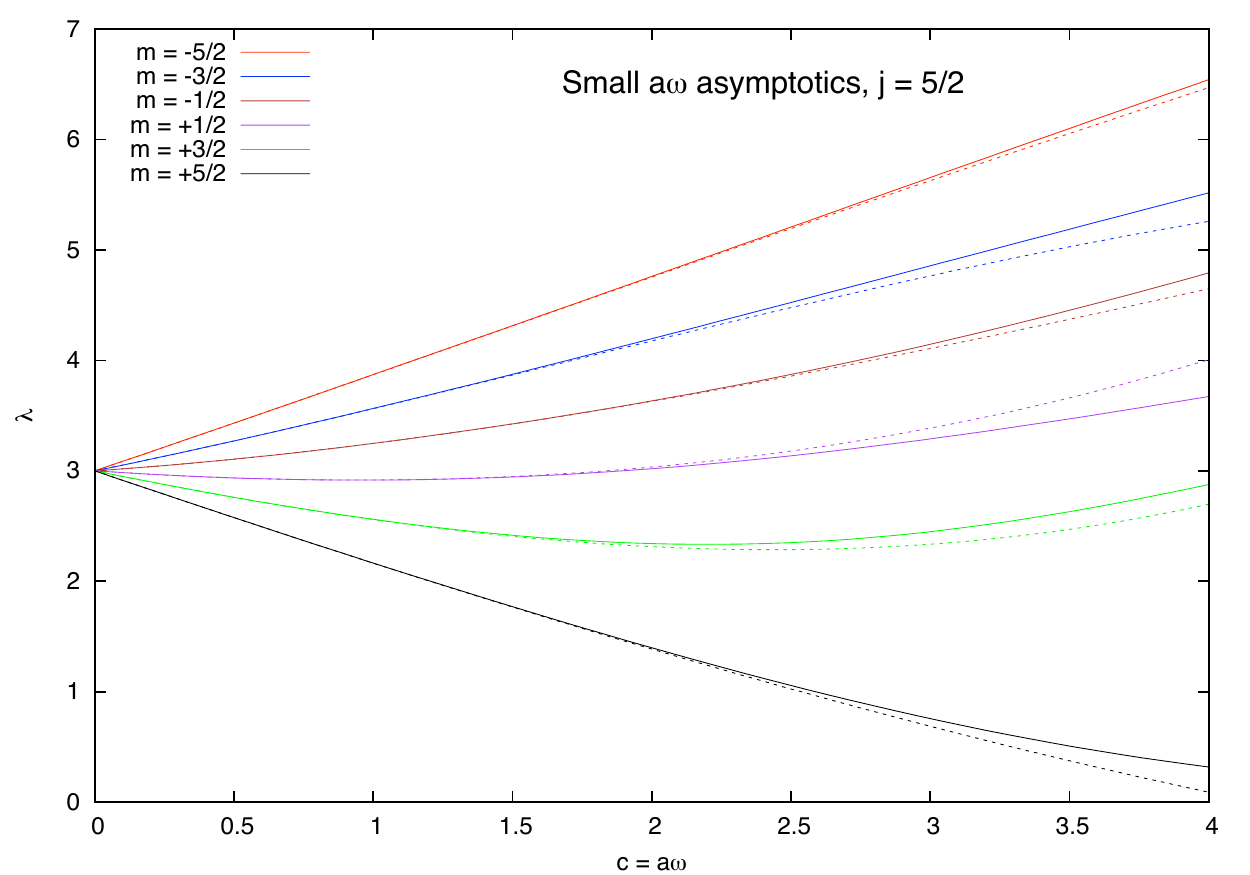}
\includegraphics[width=8cm]{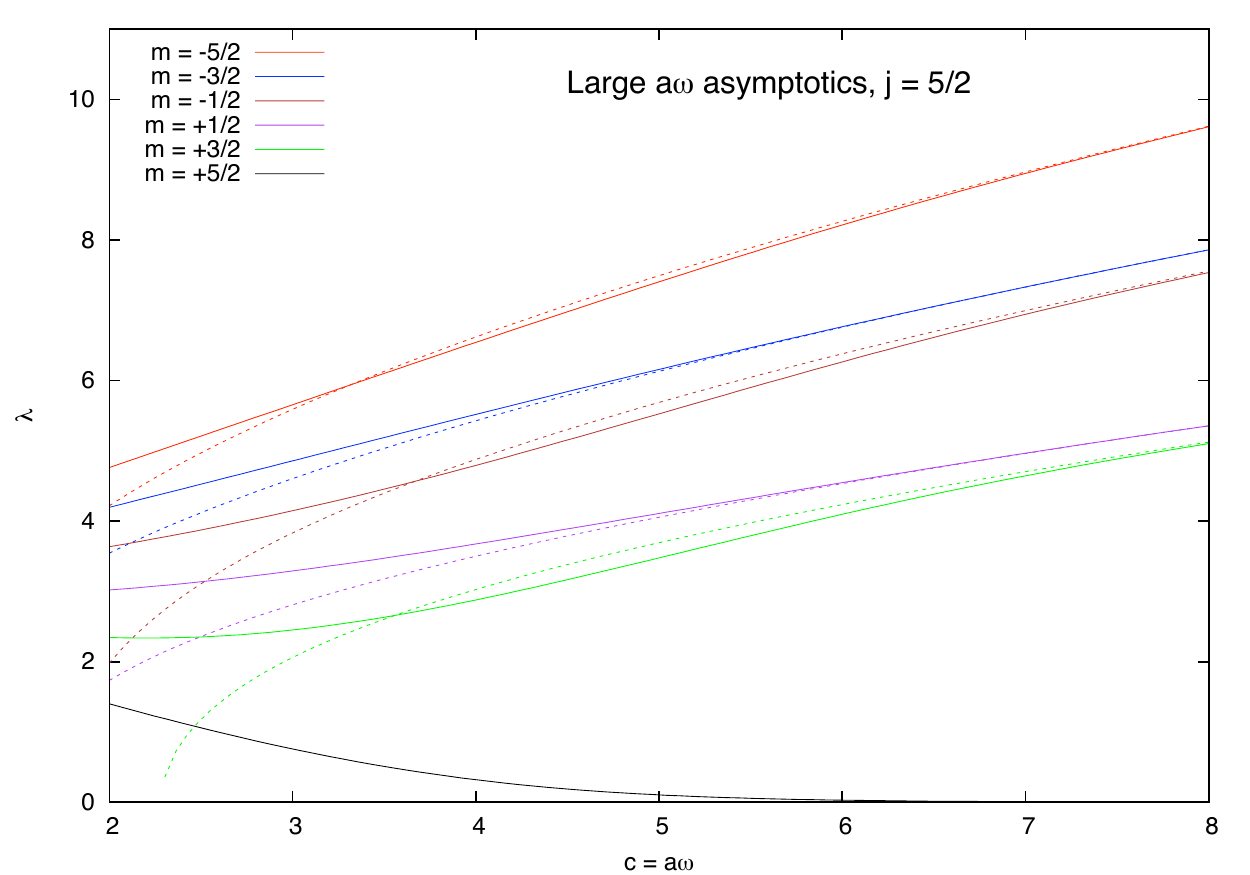}
\end{center}
\caption[]{\emph{Massless Spheroidal Eigenvalues ($a \mu = 0$)}. The top plot shows the eigenvalues $\lambda$ for modes $j = 1/2 \ldots 11/2$, $m=-j\ldots j$ in the parameter range $a\omega = 0 \ldots 8$ and $a \mu =0$.  
For positive $a \omega$, the $m = +j$ eigenvalue is smallest; it tends to zero as $a \omega \rightarrow \infty$. The massless spectrum ($a\mu = 0$) is symmetric under $\lambda \rightarrow -\lambda$. The lower plots compares numerical results [solid] with approximations [dotted] valid in the small-$a \omega$ (left) and large-$a\omega$ limits, given by Eq.~(\ref{eq:small-asymptotics}) and Eq.~(\ref{eq:large-asymptotics}).} 
\label{fig-eigenvalues-mu0}
\end{figure}

Figure \ref{fig-eigenvalues-mu0} shows eigenvalues for the massless case, $\mu = 0$, for the modes $j = 1/2, \ldots, 11/2$ and parity $\Par = +1$. The eigenvalues are split on azimuthal number $m$. For positive $a \omega$ the lowest-lying eigenvalue is the $m = j$ mode, which tends to zero as $a\omega \rightarrow \infty$. For negative $a \omega$, the symmetries (\ref{eig-sym}) mean that the spectrum looks the same but the ordering of eigenvalues in $m$ is reversed (i.e. the $m = -j$ mode is lowest-lying).  It also follows from (\ref{eig-sym}) that, in the massless case $\mu = 0$, the eigenvalues of negative parity $\Par = -1$ are found by inversion, $\lambda \rightarrow -\lambda$.

The lower plots in Fig. \ref{fig-eigenvalues-mu0} compare numerically-determined eigenvalues against asymptotic results of Sec.~\ref{sec-asymptotics}. In the small-$a \omega$ regime, we use expansion (\ref{eq:small-asymptotics}), taken to third order. In the large-$a \omega$ regime, we compare against the asymptotics (\ref{eq:large-asymptotics}). Note that there does not seem to be a satisfactory asymptote at large-$a\omega$ for the eigenvalue that approaches zero (e.g. $m=j$ for $a\omega > 0$).

\begin{figure}
\begin{center}
\includegraphics[width=12cm]{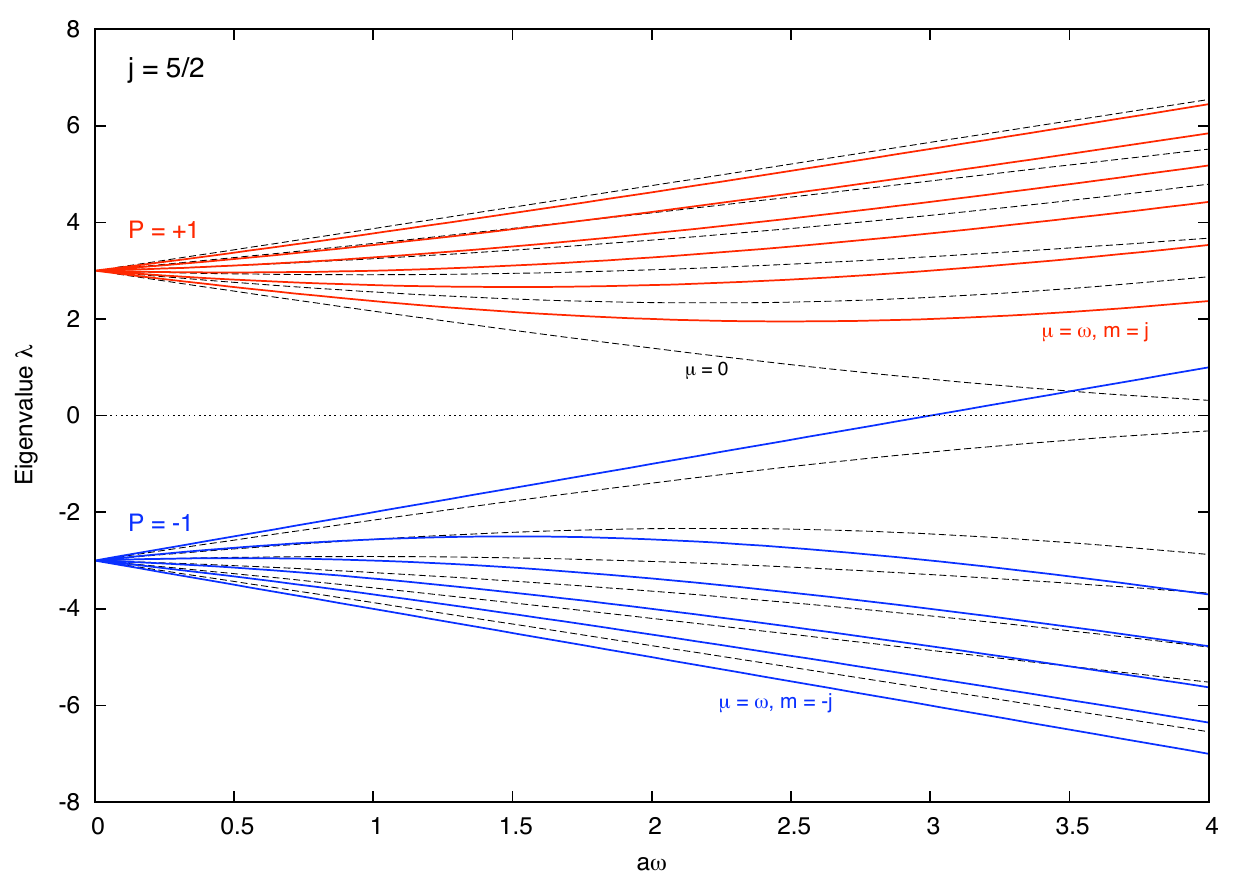}
\end{center}
\caption[]{\emph{Eigenvalues for $j = 5/2$, $m=-j \ldots j$ and $\Par = \pm 1$}. The solid lines show the $\mu = \omega$ eigenvalues (Eq. \ref{lam-exact}) for positive (red) and negative (blue) parities $\Par$. The dotted black lines show the massless eigenvalues $\mu = 0$, also shown in Fig. \ref{fig-eigenvalues-mu0}. The eigenvalue of the $a\mu = a\omega$, $m = j = 5/2$, $\Par = -1$ mode crosses the axis at $a \omega = 3$ (i.e. $\lambda_{5/2, 5/2, -1}^{(3, 3)} = 0).$}
\label{fig-eigenvalues-exact}
\end{figure}

Figure \ref{fig-eigenvalues-exact} compares the eigenvalue spectrum for the case $a\omega = a\mu$ (Eq.~(\ref{lam-exact})) with the massless ($a\mu = 0$) spectrum. The $j=5/2$ modes are shown here; other modes follow a similar pattern. Observe that the $m=j$, $P=-1$ eigenvalue crosses the $\lambda = 0$ line at high $a\omega$, but no other eigenvalue changes sign. This can be understood by examining the exact result, Eq.~(\ref{lam-exact}). For $\Par = -1$, $m = - j$ we obtain $\lambda = -(j+1/2) + a\omega$ and hence $\lambda = 0$ when $a \omega = j+1/2$. Likewise, the $m = +j$ eigenvalue is zero when $a \omega = - (j+1/2)$. For all other states, the term under the square root in (\ref{lam-exact}) is non-zero for all $a \omega \ge 0$, and hence the eigenvalue does not change sign.

\begin{figure}
\begin{center}
\includegraphics[width=11cm]{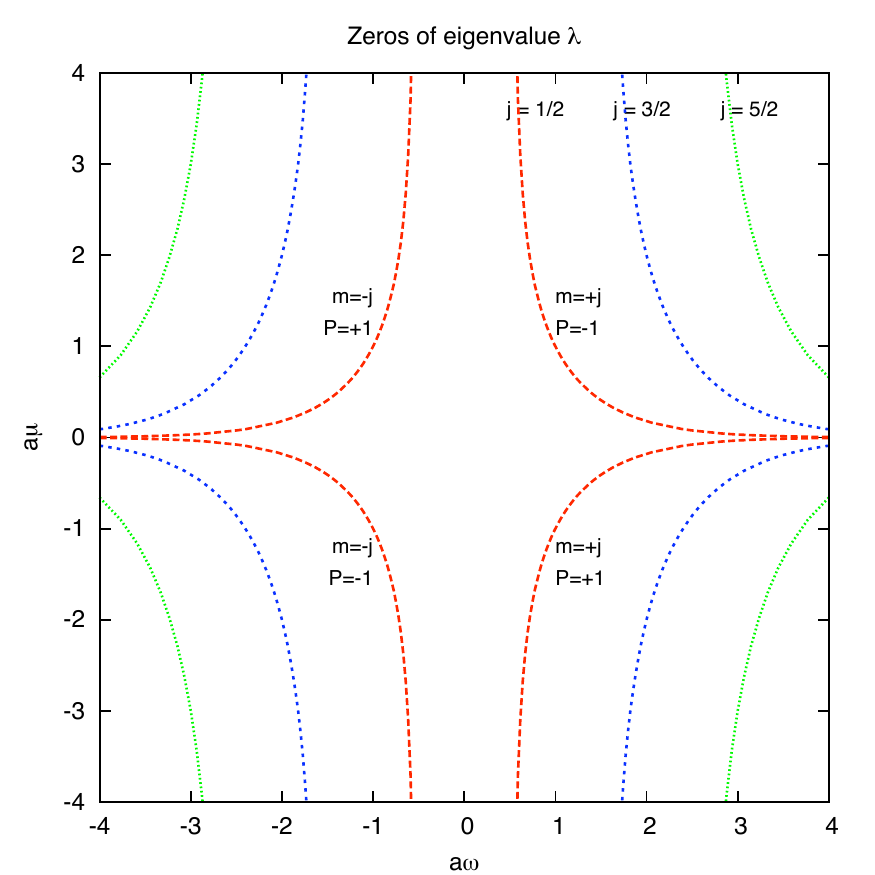}
\end{center}
\caption[]{\emph{Contours in the $a\omega$-$a\mu$ plane along which $\lambda = 0$}. The three lines show maximally co- and counter-rotating modes with $j=1/2$, $j=3/2$ and $j=5/2$ (red, blue, green). The plot illustrates the symmetries of the eigenvalue spectrum, given in Eq.~(\ref{eig-sym}).}
\label{fig-zeros}
\end{figure}

The zeros of the eigenvalue $\lambda$ are worth some further consideration. 
Figure \ref{fig-zeros} shows that, for each $j$, the $m=j$, $P=-1$ mode has zero eigenvalue along a line in the quadrant $a\omega > 0$, $a \mu > 0$. We find no evidence to suggest that the eigenvalue of any other $m$, $\Par$ mode passes through zero in this quadrant. The zeros in the other quadrants follow immediately from the symmetries of the spectrum (Eq.~\ref{eig-sym}).

In Appendix \ref{appendix-tables} we present some tables of numerically-determined eigenvalues, for  the $j=1/2$ and $j=3/2$ modes in the range $0 < a\omega < 1$, $0 < \mu / \omega < 1$. Our intention is to provide a resource for checking and validating future studies.

\subsection{Eigenfunctions}
\begin{figure}
\begin{center}
\includegraphics[width=9cm]{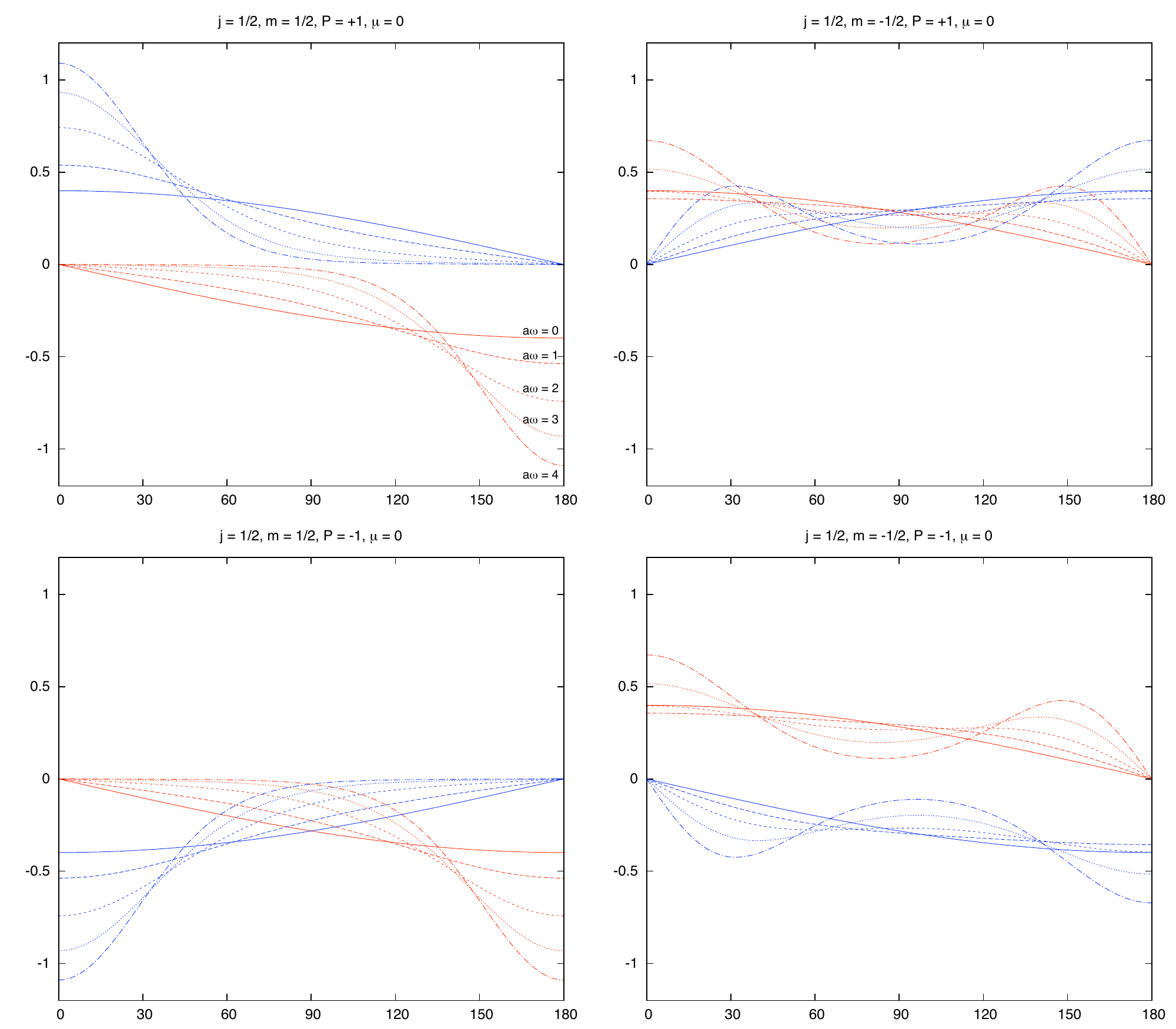} 
\includegraphics[width=16cm]{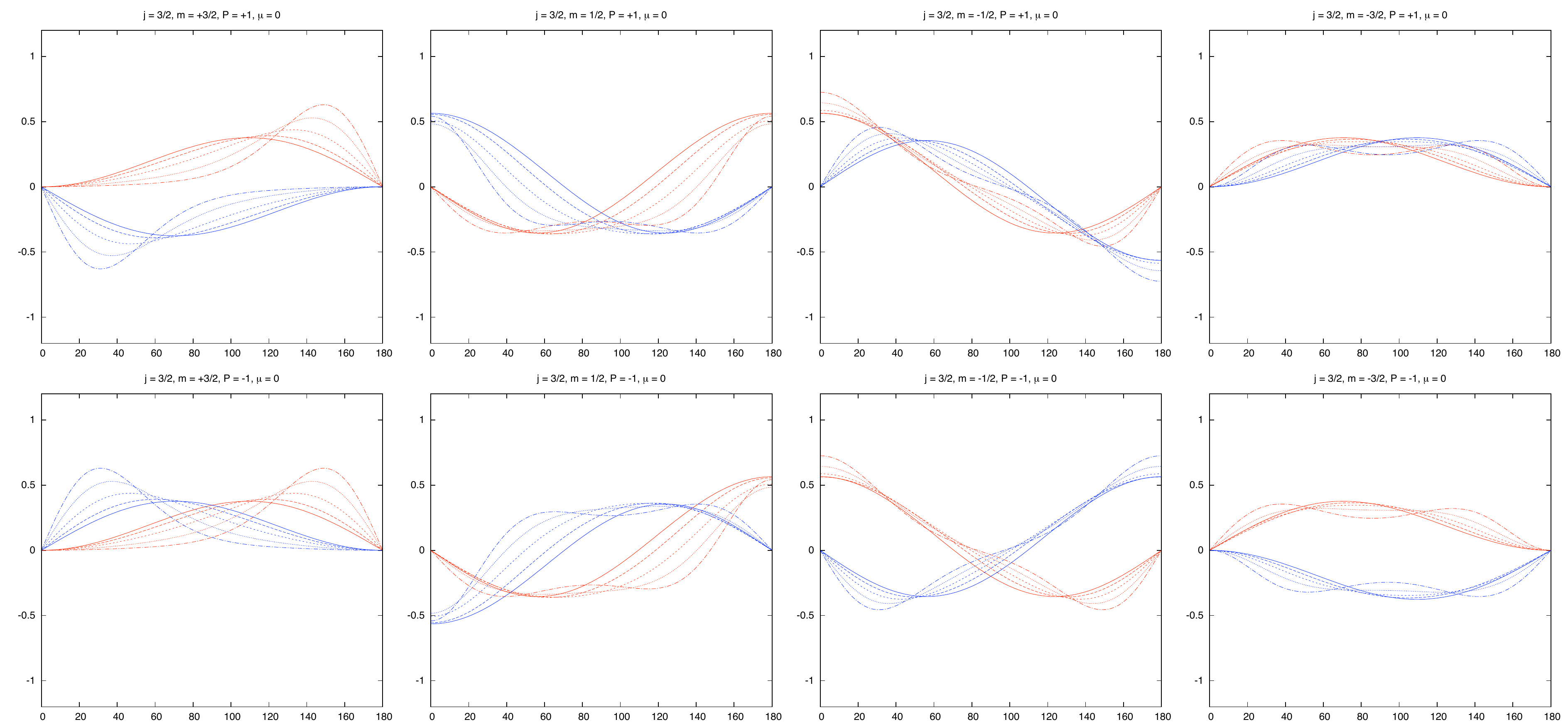}
\end{center}
\caption[]{\emph{Angular solutions $S_+(\theta)$ [red] and $S_-(\theta)$ [blue] for $a \mu = 0$: the spin-half spheroidal harmonics}. The top four plots show the $j=1/2$ modes, and the bottom eight plots show the $j=3/2$ modes, as a function of polar angle $\theta$. The co-rotating modes $m > 0$ are on the left, and the counter-rotating modes $m < 0$ are on the right. Plots come in pairs: one for positive parity (upper) and negative parity (lower), $\Par = \pm 1$. Note the symmetry under $\theta \rightarrow \pi - \theta$. The solid line shows the solution for $a \omega = 0$. The broken lines show the solutions for $a \omega = 1.0$, $2.0$, $3.0$ and $4.0$. }
\label{fig-eigenvectors-r0}
\end{figure}

\begin{figure}
\begin{center}
\includegraphics[width=9cm]{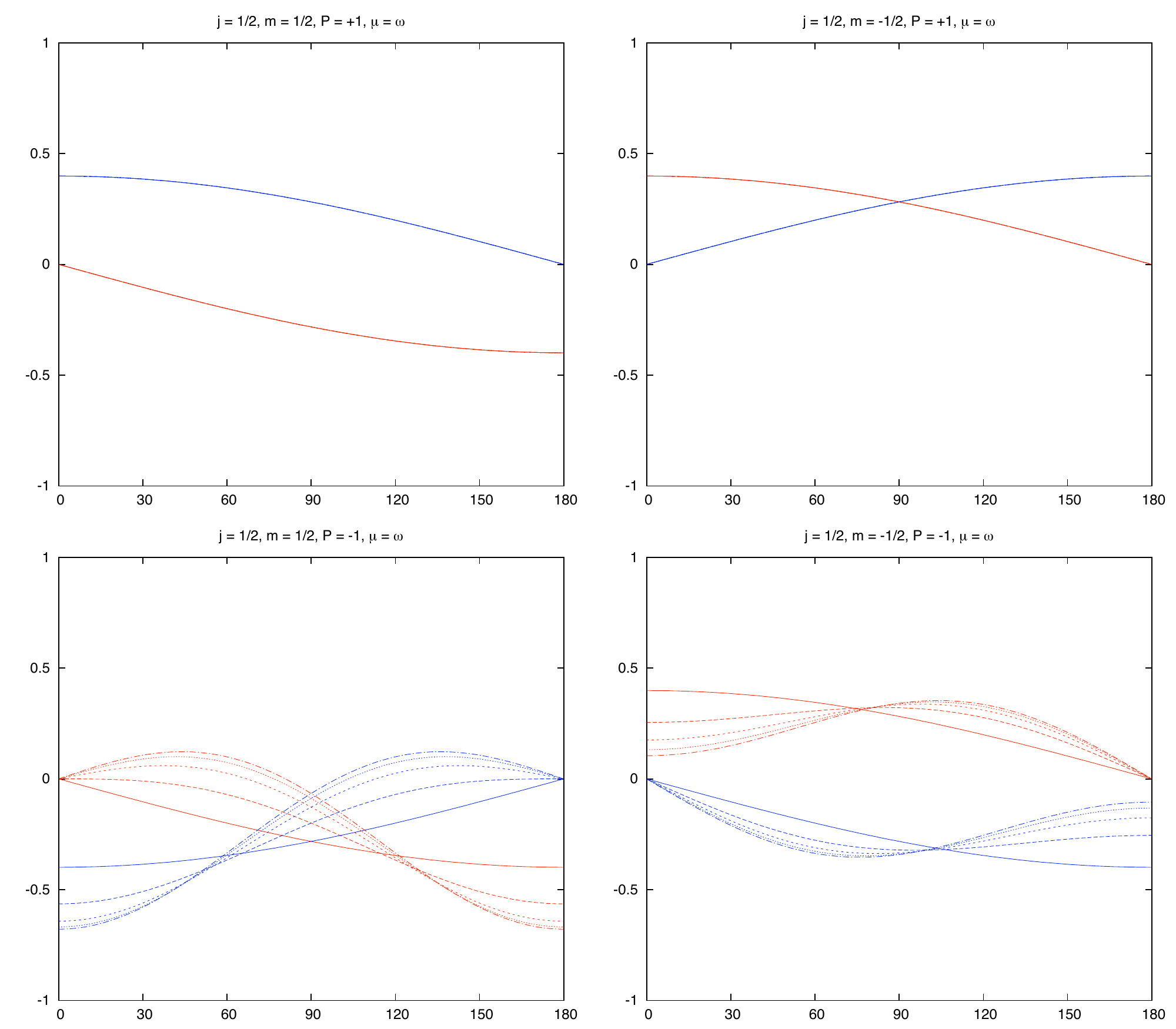} 
\includegraphics[width=16cm]{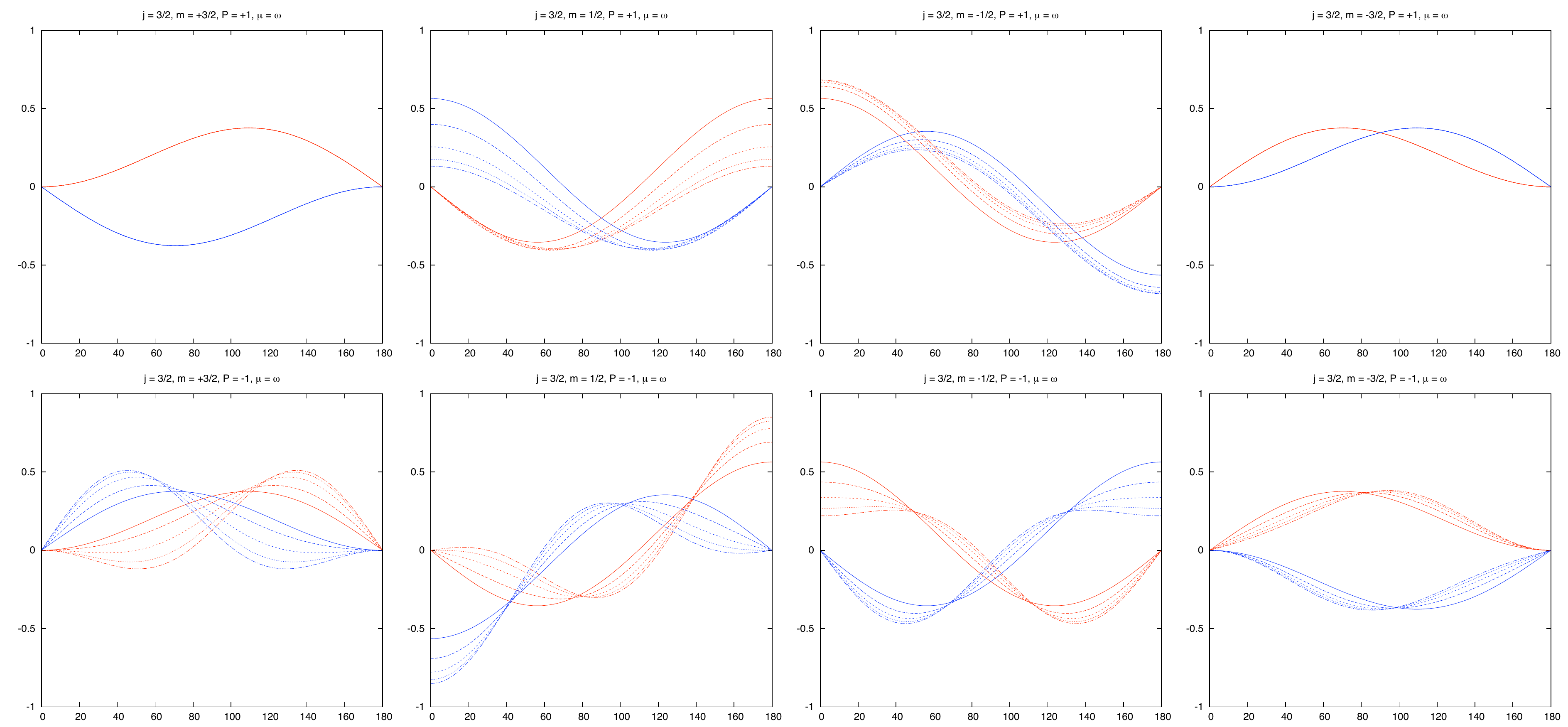}
\end{center}
\caption[]{\emph{Angular solutions $S_+(\theta)$ [red] and $S_-(\theta)$ [blue] for $a\omega = +a \mu$}. The top four plots show the $j=1/2$ modes, and the bottom eight plots show the $j=3/2$ modes, as a function of polar angle $\theta$. The co-rotating modes $m > 0$ are on the left, and the counter-rotating modes $m < 0$ are on the right. Plots come in pairs: one for positive parity (upper) and negative parity (lower), $\Par = \pm 1$. The solid line shows the solution for $a\omega = 0$. The broken lines show the solutions for $a\omega = 1.0$, $2.0$, $3.0$ and $4.0$. }
\label{fig-eigenvectors-r+1}
\end{figure}

Figures \ref{fig-eigenvectors-r0} and \ref{fig-eigenvectors-r+1} show the eigenfunctions $S_+(\theta)$ (red) and $S_-(\theta)$ (blue) of the $j = 1/2$ and $j = 3/2$ modes, for a range of $a\omega \ge 0$ and $a\mu \ge 0$. 
Figure \ref{fig-eigenvectors-r0} shows the massless spheroidal harmonics ($a \mu = 0$). The solid line represents the spherical ($a=0$) harmonics, and the broken lines represent the spheroidal harmonics with $a\omega = 1, 2, 3$ and $4$. As the rotational coupling increases, the eigenfunctions show a tendency to increase in magnitude near the poles ($\theta \sim 0^\circ, \theta \sim 180^\circ$), and decrease in magnitude around the equatorial plane ($\theta = 90^\circ$). 

Figure \ref{fig-eigenvectors-r+1} shows $j=1/2$, $j=3/2$ eigenfunctions for the special case $\omega =  \mu$, in the range $a\omega = 0 \ldots 4$. In this case, the eigenfunctions are known in closed form (\ref{Smu-pos-defn}). The plot makes it clear that the eigenfunctions of the $\Par = +1$, $m = \pm j$, $a\mu = a\omega$ mode do not depend on $a \omega$. For the other modes, we again see a general `enhancement' towards the poles. 

The eigenfunctions plotted in Figs \ref{fig-eigenvectors-r0} and \ref{fig-eigenvectors-r+1} exhibit the symmetries stated in Eq. (\ref{S-sym1})--(\ref{S-sym2}). For example, the reflection symmetry $S_+(\theta) = \Par (-1)^{j + m} S_-(\pi - \theta)$ is obvious. In the massless case (Fig.\ref{fig-eigenvectors-r0}) there is also a clear symmetry under parity change $\Par = - \Par$; whereas in the general case, this only holds if we simultaneously flip the sign of the mass ($\mu \rightarrow - \mu$). Solutions for arbitrary $a \omega$, $a \mu$ can be found by applying the symmetries (\ref{S-sym1})--(\ref{S-sym3}) to the solutions in the first quadrant.

\begin{figure}
\begin{center}
\includegraphics[width=16cm]{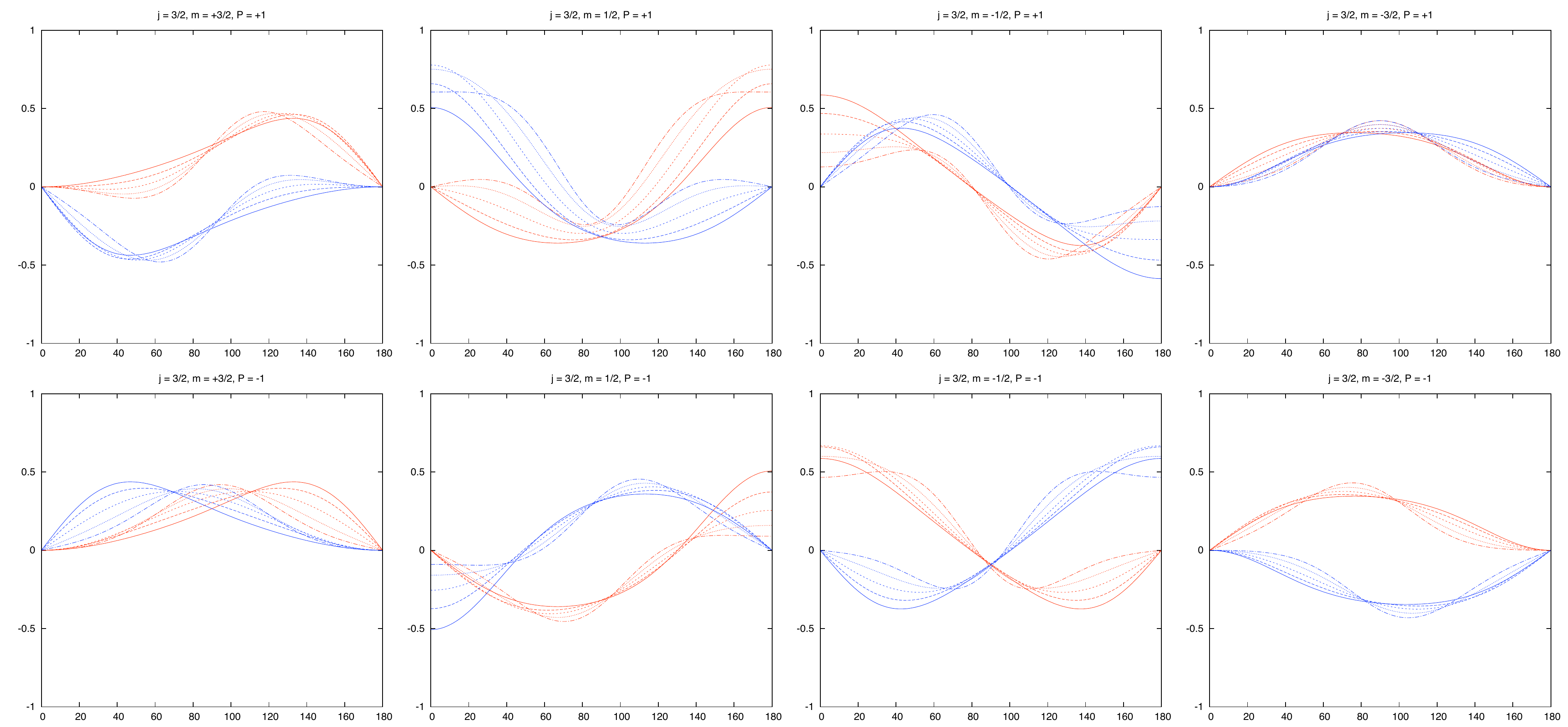}
\end{center}
\caption[]{\emph{Angular solutions for $j = 3/2$ and $a \omega = 2.0$ and $a\mu = 0, 1.0, 2.0, 3.0, 4.0$}. The solid line shows the massless solution, $a\mu = 0$. The broken lines show the solutions for $a\omega = 1.0 \ldots 4.0$.}
\label{fig-eigenvectors-amu}
\end{figure}

Figure \ref{fig-eigenvectors-amu} shows the influence of field mass on the shape of the eigenfunctions. Here, we plot the eigenfunctions of the $j=1/2$, $j=3/2$ modes, for mass-to-frequency ratios $\mu / \omega$ from zero to two. In general, it would seem that mass seems to lead to additional structure in the eigensolutions around the equatorial plane. However, we defer any attempt at physical interpretation to a future in-depth study.


\section{Conclusion\label{sec-conclusion}}
We have described a technique for the determination of eigenvalues and eigenfunctions of the angular equation for the massive Dirac field on a rotating black-hole background. By carrying out a spectral decomposition of the angular eigenfunctions, we obtained a three-term recurrence relation. Exact eigenvalues can be easily and quickly obtained by numerically solving the three-term relation (\ref{recurr}); alternatively, they may be estimated from series expansion (\ref{eq:small-asymptotics}). 
We have described two alternative methods to obtain the eigenfunctions, including an alternative set of recurrence relations that were derived in an attempt to correct one of the previous attempts to obtain a solution for the eigenvalues~\cite{Batic-Nowakowski-2008}. We have tabulated eigenvalues for a range of values of the spin of the central black hole and the mass of the fermion field (Tables~\ref{table-j1-P1}--\ref{table-j3-m1-P-1}), and derived fits which reproduce the numerical results to high precision (Tables~\ref{fit-j1-P1}--\ref{fit-j3-m1-P-1}).

The motivation for this work was the need for angular eigenvalues as input for two separate studies --- an investigation of bound states in the massless Kerr-Newman background~\cite{Gair-thesis} and a study of  the spectrum of quasi-bound states in the vicinity of a small black hole~\cite{Lasenby-2005, Dolan-2007}. The solution to the radial equation for these two problems will be presented elsewhere, together with an estimate of the error arising from using fitting function Eq.~(\ref{lamfit}) in place of exact angular eigenvalues. 

We hope this work will also be of benefit to other studies, for example, investigations into (i) the effect of field mass on fermionic quasi-normal ringing of black holes;  (ii) the effect of particle mass on the Hawking radiation emission spectrum for temperatures close to the mass of the particle species; and (iii) the interaction of brane-localised fermions with rotating higher-dimensional black holes.

\newpage
\newpage
\newpage
\newpage
\newpage
\newpage
\newpage
\newpage
\newpage

\begin{acknowledgments}
SD would like to thank Anthony Lasenby and Chris Doran for helpful discussions and geometric insight, and Adrian Ottewill and Marc Casals, and the Irish Research Council for Science, Engineering and Technology (IRCSET) for financial support. JG would like to thank Donald Lynden-Bell for useful discussions. JG's work is supported by the Royal Society.
\end{acknowledgments}

\bibliography{chandrapage_sdjg}
\bibliographystyle{apsrev}

\appendix
\section{Clebsch-Gordan Coefficients\label{appendix-clebsch}}
In this section we give explicit forms for the Clebsch-Gordan coefficients that appear in integrals (\ref{bigC-defn}) and (\ref{bigD-defn}), and present simplified expressions for $C_{kk^\prime}^{(1)}$ and $D_{kk^\prime}^{(1)}$. The relevant coefficients are
\begin{eqnarray}
\left< k-1, \, 1 \, m \, 0 \, | \, k \, m \right> &=& \sqrt{ \frac{2 [k^2 - m^2]}{(2k) (2 k-1) } }  , \\
\left< k, \, 1 \, m \, 0 \, | \, k \, m \right> &=&  \frac{m}{ \sqrt{k(k+1)} }  ,  \\
\left< k+1, \, 1 \, m \, 0 \, | \, k \, m \right> &=&  -\sqrt{ \frac{2 [(k+1)^2 - m^2]}{(2k+3)(2k+2)}} ,
\end{eqnarray}
and
\begin{eqnarray}
\left< k-1, \, 1 \, \tfrac{1}{2} \, 0 \, | \, k \, \tfrac{1}{2} \right> &=& \frac{1}{2} \sqrt{ \frac{2k+1}{k} } \\
\left< k,  \, 1 \, \tfrac{1}{2} \, 0 \, | \, k \, \tfrac{1}{2} \right> &=&  \frac{1}{2} \sqrt{\frac{1}{k(k+1)}}   \\
\left< k+1, \, 1 \, \tfrac{1}{2} \, 0 \, | \, k \, \tfrac{1}{2} \right> &=& -\frac{1}{2} \sqrt{\frac{2k+1}{k+1}}  .
\end{eqnarray}
and
\begin{eqnarray}
\left< k-1, \, 1 \, -\tfrac{1}{2} \, 1 \, | \, k \, \tfrac{1}{2} \right> &=&  \sqrt{ \frac{ 2k+1 }{ 8k } } \\
\left< k,     \, 1 \, -\tfrac{1}{2} \, 1 \, | \, k \, \tfrac{1}{2} \right> &=&  -\frac{2k+1}{\sqrt{8k (k+1)}}  \\
\left< k+1, \, 1 \, -\tfrac{1}{2} \, 1 \, | \, k \, \tfrac{1}{2} \right> &=&  \sqrt{ \frac{ 2k+1 }{ 8(k+1) } } 
\end{eqnarray}
Hence the integrals (\ref{bigC-defn}) and (\ref{bigD-defn}) are 
\begin{eqnarray}
C_{kk^\prime}^{(1)} &=& \left\{ \begin{array}{ll}
 \sqrt{k^2 - m^2} / (2k) &  k^\prime = k - 1  \\  
 m / [2k (k+1)] &  k^\prime = k \\
 \sqrt{(k+1)^2 - m^2} / [2(k+1)] \quad \quad &  k^\prime = k + 1 \\
 0 & |k - k^\prime| > 1 .
\end{array}    \right. \\
\Par D_{kk^\prime}^{(1)} &=&  \left\{  \begin{array}{ll}  
\sqrt{k^2 - m^2} / (2k) & k^\prime = k - 1  \\  
- m (k+1/2) / [k(k+1)] & k^\prime = k \\
- \sqrt{(k+1)^2 - m^2} / [2 (k+1) ]  \quad \quad &  k^\prime = k + 1 \\
 0 & |k - k^\prime| > 1 .
\end{array} \right.
\end{eqnarray}

\section{Tables of Eigenvalues\label{appendix-tables}}
Numerical eigenvalues for the $j=1/2$ and $j=/2$ modes are tabulated in Tables \ref{table-j1-P1}--\ref{table-j3-m1-P-1}.  The parameter range is the same as in \cite{Chakrabarti-1984}, $a\sig = 0 \ldots 1$ and $r = \mu / \sig = 0 \ldots 1$. The eigenvalues are accurate to six decimal places, and have been checked using the method of SFC \cite{Suffern-1983}. They disagree with the eigenvalues presented in Chakrabarti \cite{Chakrabarti-1984}, Tables I, IIa, IIb and III. It is possible to find simple polynomial fits that reproduce these eigenvalues with high precision. We take an ansatz of the form, and fit it in the range $a\om \in [-1,0]$, $r=\mu/\om \in [0,1]$
\begin{equation}
\lambda = \lambda_0 + \sum_{n=1}^{n=4} \sum_{k=0}^{k=4} f_{nk} (a\om)^n (1+\mu/\om)^k .
\label{lamfit}
\end{equation}
The coefficients $f_{nk}$ that reproduce the data in Tables~\ref{table-j1-P1}--\ref{table-j3-m1-P-1} are tabulated in Tables~\ref{fit-j1-P1}--\ref{fit-j3-m1-P-1}.

\begin{table} 
\begin{tabular}{| l | l l l l l |}
\hline
$P = +1$& & & $a \omega$ && \\
$R = \mu/\omega$ \hspace{1cm}&        0.1\hspace{4cm}&             0.2\hspace{2.5cm}&             0.3\hspace{2.5cm}&             0.4\hspace{2.5cm}&             0.5\hspace{2.5cm}\\
\hline
0.0\quad\quad\quad&  0.934097   \quad\quad\quad&   0.869818    \quad\quad\quad&    0.807306   \quad\quad\quad&   0.746712    \quad\quad\quad&   0.688186    \quad\quad\quad\\
      &	 1.067385 & 1.136116      &   1.206065    &   1.277108    &  1.349131    \\
0.2&  0.941098 & 0.884521      &   0.830461    &   0.779115    &  0.730674     \\
      & 1.061036 &  1.124021     &    1.188780   &    1.255145   &   1.322958    \\
0.4&  0.948159 & 0.899463      &  0.854152     &   0.812462    &  0.774612     \\
      & 1.054745 &  1.112158     &   1.172009    &    1.234081   &   1.298167    \\
0.6&  0.955278  &  0.914640     &  0.878364     &  0.846712    &   0.819912    \\
      & 1.048514 &   1.100528    &   1.155756    &    1.213920   &   1.274757    \\
0.8&  0.962457 &   0.930048    &  0.903081     &  0.881822     &   0.866479    \\
      & 1.042342 &   1.089133    &   1.140022    &    1.194663   &    1.252724   \\
1.0&  0.969694 &   0.945683    &  0.928286     &  0.917745     &  0.914214     \\
      & 1.036229 &    1.077973   &   1.124808    &   1.176305    &   1.232051   \\
&&&&&\\
        &0.6&             0.7&             0.8&             0.9&             1.0\\
        \hline
0.0& 0.631876  &    0.577928   &  0.526476     &    0.477646   &   0.431544   \\
      & 1.422021 &     1.495674  &   1.569992    &     1.644879  &    1.720244   \\
0.2& 0.685319  &    0.643219   &  0.604522     &    0.569349   &   0.537786   \\
      & 1.392071 &     1.462344  &   1.533646    &     1.605854  &    1.678854   \\
0.4& 0.740800  &    0.711195   &  0.685925     &    0.665076    &  0.648682   \\
       &1.364073 &     1.431621  &  1.500643     &     1.570983  &    1.642500   \\
0.6& 0.798151  &    0.781565   &  0.770229     &    0.764154   &   0.763286    \\
      & 1.338022 &     1.403483  &   1.470931    &     1.540169  &    1.611021   \\
0.8& 0.857194  &    0.854030   &  0.856969     &    0.865911   &    0.880683   \\
      & 1.313893 &     1.377878  &   1.444410     &    1.513243   &   1.584153   \\
1.0& 0.917745  &    0.928286   &  0.945683     &    0.969694   &   1.000000   \\
      & 1.291647 &     1.354724  &   1.420937    &     1.489975  &    1.561553   \\
\hline
\end{tabular}
\caption{Eigenvalues $\lambda$ for $j = 1/2, m = \pm 1/2$ and $P = +1$. In each row, the top line is the $m=+1/2$ eigenvalue and the lower line is the $m=-1/2$ value. Here, $R$ is the ratio between mass $\mu$ and frequency $\omega$. }
\label{table-j1-P1}
\end{table}

\begin{table} 
\begin{tabular}{| l | l l l l l |}
\hline
$P = -1$& & & $a \omega$ && \\
$R = \mu / \omega$\hspace{1cm}&        0.1\hspace{4cm}&             0.2\hspace{2.5cm}&             0.3\hspace{2.5cm}&             0.4\hspace{2.5cm}&             0.5\hspace{2.5cm}\\
\hline
0.0\quad\quad\quad&  -0.934097   \quad\quad\quad&   -0.869818    \quad\quad\quad&   -0.807306    \quad\quad\quad&   -0.746712    \quad\quad\quad&   -0.688186    \quad\quad\quad\\
      &	-1.067385 &     -1.136116  &    -1.206065    &   -1.277108   &    -1.349131   \\
0.2& -0.927156  &    -0.855357   &   -0.784701    &   -0.715290    &   -0.647230    \\
      & -1.073792 &     -1.148441  &    -1.223859   &   -1.299963   &     -1.376672   \\
0.4& -0.920276   &   -0.841142    &  -0.762657    &   -0.684881    &   -0.607877    \\
      & -1.080258 &     -1.160994  &   -1.242157   &    -1.323697   &    -1.405563   \\
0.6& -0.913456  &    -0.827176   &   -0.741186    &   -0.655515     &   -0.570190    \\
      & -1.086781 &     -1.173773  &    -1.260952   &    -1.348294   &    -1.435774  \\
0.8& -0.906697  &    -0.813461   &   -0.720298    &   -0.627215    &   -0.534219    \\
      & -1.093362 &    -1.186776   &    -1.280236   &    -1.373736   &    -1.467268   \\
1.0& -0.900000  &   -0.800000    &   -0.700000    &   -0.600000    &   -0.500000    \\
      & -1.100000 &    -1.200000   &   -1.300000    &    -1.400000   &    -1.500000   \\
&&&&&\\
        &0.6&             0.7&             0.8&             0.9&             1.0\\
        \hline
0.0& -0.631876  &   -0.577928    &  -0.526476     &   -0.477646    &   -0.431544    \\
      & -1.422021 &   -1.495674    &   -1.569992    &   -1.644879    &    -1.720244   \\
0.2& -0.580627  &   -0.515589    &  -0.452221     &   -0.390625    &   -0.330897    \\
      & -1.453910 &    -1.531603   &   -1.609681    &   -1.688073    &    -1.766714   \\
0.4&  -0.531711 &  -0.456447     &  -0.382152     &   -0.308894    &   -0.236738     \\
       & -1.487709 &  -1.570089     &  -1.652658    &    -1.735371   &    -1.818185   \\
0.6& -0.485243  &  -0.400704     &  -0.316603     &   -0.232973     &  -0.149846     \\
      & -1.523370 &   -1.611058    &   -1.698816    &    -1.786622   &   -1.874453    \\
0.8& -0.441318  &  -0.348520     &  -0.255831     &   -0.163261    &   -0.070817    \\
      & -1.560827 &   -1.654406    &   -1.747998    &    -1.841598   &   -1.935199    \\
1.0& -0.400000  &  -0.300000     &  -0.200000     &   -0.100000    &   0.000000    \\
      & -1.600000 &   -1.700000    &   -1.800000    &    -1.900000    &  -2.000000      \\
\hline
\end{tabular}
\caption{Eigenvalues $\lambda$ for $j = 1/2, m = \pm 1/2$ and $P = -1$. In each row, the top line is the $m=+1/2$ eigenvalue and the lower line is the $m=-1/2$ value. Here, $R$ is the ratio between mass $\mu$ and frequency $\omega$. }
\label{table-j1-P-1}
\end{table}

\begin{table} 
\begin{tabular}{| l | l l l l l |}
\hline
$P = +1$&&& $a \omega$ && \\
$R=\mu/\omega$  &0.1&           0.2&            0.3&            0.4&            0.5\\
\hline
0.0\quad\quad\quad& 1.920329  \quad\quad\quad&   1.841350   \quad\quad\quad&  1.763119    \quad\quad\quad&  1.685695    \quad\quad\quad&   1.609143   \quad\quad\quad \\
      & 2.080312 & 2.161215    &  2.242663   &  2.324614   &   2.407027   \\
0.2& 1.924473 & 1.849942   &  1.776486    &  1.704190    &  1.633144    \\
      & 2.076449 &  2.153751   & 2.231842    &  2.310661   & 2.390152     \\
0.4& 1.928643 & 1.858641   &  1.790098    &  1.723128    &  1.657851    \\
      & 2.072612 & 2.146386   &  2.221237    & 2.297085    &  2.373856    \\
0.6& 1.932840&  1.867446   &  1.803952    &   1.742503   &   1.683251   \\
      & 2.068799 &  2.139119   &  2.210849   &  2.283888    &  2.358139   \\
0.8& 1.937063 &  1.876356  &  1.818048    &  1.762311    &  1.709329    \\
      & 2.065012 & 2.131951 &  2.200679     &  2.271067    &  2.342997    \\
1.0& 1.941311  &  1.885372  &  1.832381    &  1.782542    &  1.736068    \\
      & 2.061250 &  2.124881   &  2.190725    &  2.258623    &  2.328427    \\
&&&&&\\
        &0.6&           0.7&            0.8&            0.9&            1.0\\
        \hline
0.0& 1.533529 &  1.458925    &  1.385410    &  1.313063    &  1.241970    \\
      & 2.489862 &  2.573085   &   2.656660   &   2.740555   &   2.824737   \\
0.2& 1.563445 &  1.495195    &  1.428503   &   1.363483   &   1.300256   \\
      & 2.470262 &  2.550942   &   2.632146   &   2.713831   &   2.795954   \\
0.4& 1.594395 &  1.532895    &  1.473492    &  1.416335    &  1.361576    \\
      & 2.451480 &  2.529894   &   2.609036    &  2.688852   &   2.769289   \\
0.6& 1.626354 &  1.571979    &  1.520298    &  1.471488    &  1.425727   \\
      & 2.433514 &  2.509932  &    2.587316   &   2.665597   &    2.744709 \\
0.8& 1.659291 &  1.612391    &  1.568825   &   1.528789   &   1.492475   \\
      & 2.416357&  2.491044   &   2.566964   &   2.644029    &   2.722158   \\
1.0& 1.693171 &  1.654066    &  1.618962    &   1.588061   &  1.561553   \\
      & 2.400000 &  2.473214   &  2.547950    &  2.624100    &   2.701562   \\
\hline
\end{tabular}
\caption{Eigenvalues for $j = 3/2, m = \pm 3/2$ and $P = +1$.}
\label{table-j3-m3-P1}
\end{table}

\begin{table} 
\begin{tabular}{| l | l l l l l |}
\hline
$P = -1$&&& $a \omega$ && \\
$R=\mu/\omega$        &0.1&           0.2&            0.3&            0.4&            0.5\\
\hline
0.0\quad\quad\quad& -1.920329 \quad\quad\quad&   -1.841350   \quad\quad\quad&  -1.763119    \quad\quad\quad&   -1.685695   \quad\quad\quad&  -1.609143    \quad\quad\quad\\
        &-2.080312 & -2.161215     &   -2.242663   &  -2.324614    &  -2.407027    \\
0.2& -1.916210  & -1.832864     &   -1.749998   &  -1.667649    &   -1.585858   \\
      & -2.084199 & -2.168777    &    -2.253701   &  -2.338945    &  -2.424480    \\
0.4& -1.912118  & -1.824486     &   -1.737125   &  -1.650054    &   -1.563297   \\
      & -2.088112 & -2.176436    &    -2.264955  &   -2.353651    &  -2.442509   \\
0.6& -1.908053  & -1.816216     &  -1.724500    &  -1.632913    &  -1.541465    \\
      & -2.092050 &  -2.184194   &   -2.276423   &   -2.368731    &   -2.461110   \\
0.8& -1.904013  & -1.808054     &  -1.712125    &  -1.616228    &  -1.520366    \\
      & -2.096012 & -2.192048    &    -2.288106   &  -2.384182    &   -2.480276   \\
1.0& -1.900000  & -1.800000     &  -1.700000    &  -1.600000    &  -1.500000    \\
      & -2.100000 & -2.200000     &   -2.300000    &   -2.400000   &  -2.500000    \\
&&&&&\\
        &0.6&           0.7&            0.8&            0.9&            1.0\\
        \hline
0.0& -1.533529  &   -1.458925   &   -1.385410   &  -1.313063    &  -1.241970    \\
      & -2.489862 &   -2.573085    &  -2.656660   &   -2.740555   &  -2.824737    \\
0.2& -1.504668  &  -1.424125    &  -1.344277    &  -1.265176    &  -1.186878    \\
      & -2.510280 &   -2.596321   &   -2.682579    &  -2.769030    &  -2.855652    \\
0.4& -1.476877  &  -1.390819    &  -1.305151    &  -1.219898    &  -1.135092    \\
      & -2.531512 &   -2.620646   &   -2.709896    &  -2.799247   &   -2.888686   \\
0.6& -1.450166  & -1.359027     &  -1.268059    &  -1.177273    &  -1.086681    \\
      & -2.553551 &  -2.646048    &  -2.738593    &   -2.831179   &   -2.923800   \\
0.8&  -1.424541 & -1.328755     &  -1.233011    &  -1.137312     & -1.041660     \\
      & -2.576385  & -2.672507    &   -2.768640   &   -2.864783   &  -2.960933    \\
1.0& -1.400000  & -1.300000     &  -1.200000   &  -1.100000    &  -1.000000    \\
      & -2.600000 &  -2.700000    &   -2.800000   &  -2.900000    &  -3.000000    \\
\hline
\end{tabular}
\caption{Eigenvalues for $j = 3/2, m = \pm 3/2$ and $P = -1$.}
\label{table-j3-m3-P-1}
\end{table}

\begin{table} 
\begin{tabular}{| l | l l l l l |}
\hline
$P = +1$& & & $a \omega$ && \\
$R=\mu/\omega$  &0.1&           0.2&            0.3&            0.4&            0.5\\
\hline
0.0\quad\quad\quad& 1.974582 \quad\quad\quad&   1.951771   \quad\quad\quad&  1.931737    \quad\quad\quad&   1.914653   \quad\quad\quad&   1.900690   \quad\quad\quad\\
        & 2.027860 &  2.058002    &   2.090270   &   2.124517   &  2.160601    \\
0.20& 1.975854 &  1.954168    &   1.935077   &   1.918712   &  1.905206    \\
        & 2.026477 &  2.055157    &   2.085916   &   2.118632   &  2.153187    \\
0.40& 1.977225 &  1.956967    &   1.939328   &   1.924407   &  1.912300    \\
        & 2.025190 &  2.052692    &   2.082402   &   2.114216   &  2.148033    \\
0.60& 1.978695 &  1.960169    &   1.944501   &   1.931764  &   1.922023   \\
        & 2.023999 &  2.050604    &   2.079720   &   2.111253   &  2.145105    \\
0.80& 1.980265 &  1.963777   &    1.950603   &   1.940800   &  1.934412   \\
        & 2.022904 &  2.048890   &     2.077863 &   2.109722   &   2.144362   \\
1.00& 1.981935 &  1.967793    &   1.957641   &   1.951530   &    1.949490  \\
        & 2.021904 &  2.047548   &    2.076820  &   2.109598   &   2.145751   \\
&&&&&\\
        &0.6&           0.7&            0.8&            0.9&            1.0\\
        \hline
0.00& 1.890016 &  1.882792    &   1.879170   &  1.879284    &   1.883249   \\
        & 2.198388 &  2.237750    &   2.278567   &  2.320724    &   2.364112   \\
0.20& 1.894692 &  1.887297    &   1.883142   &  1.882341   &    1.884998  \\
        & 2.189469 &  2.227369   &    2.266780   &  2.307602    &   2.349738  \\
0.40& 1.903101 &  1.896895   &    1.893767  &   1.893790   &    1.897033  \\
        & 2.183752 &  2.221273   &    2.260500  &   2.301337   &    2.343692  \\
0.60& 1.915337 &  1.911754    &   1.911315   &  1.914052   &    1.919984  \\
        & 2.181178 &  2.219372   &    2.259589  &   2.301732   &    2.345703  \\
0.80& 1.931468 &  1.931986    &   1.935967   &  1.943401    &   1.954263   \\
        & 2.181673 &  2.221543    &    2.263861  &  2.308514    &   2.355390  \\
1.00& 1.951530 &  1.957641    &   1.967793  &   1.981935   &    2.000000  \\
        & 2.185144 &  2.227636    &   2.273085  &   2.321347   &    2.372281  \\
\hline
\end{tabular}
\caption{Eigenvalues for $j = 3/2, m = \pm 1/2$ and $P = +1$.}
\label{table-j3-m1-P1}
\end{table}

\begin{table} 
\begin{tabular}{| l | l l l l l |}
\hline
$P = -1$&&& $a \omega$ && \\
$R=\mu / \omega$        &0.1&           0.2&            0.3&            0.4&            0.5\\
\hline
0.0\quad\quad\quad& -1.974582  \quad\quad\quad&  -1.951771    \quad\quad\quad&   -1.931737   \quad\quad\quad&   -1.914653   \quad\quad\quad&   -1.900690   \quad\quad\quad\\
        & -2.027860 &   -2.058002  &  -2.090270    &   -2.124517   &   -2.160601   \\
0.2& -1.973408  &  -1.949771    &   -1.929299   &   -1.912203   &  -1.898688    \\
        & -2.029340 &  -2.061228   &   -2.095471   &   -2.131886   &  -2.170299    \\
0.4& -1.972333  &  -1.948166    &   -1.927750    &  -1.911326    &  -1.899125    \\
        & -2.030917 &  -2.064837   &   -2.101522    &  -2.140746    &  -2.182294    \\
0.6& -1.971356  &  -1.946953     &   -1.927075   &  -1.911986    &  -1.901920    \\
        & -2.032590 &  -2.068830   &   -2.108428   &   -2.151104   &  -2.196590    \\
0.8& -1.970476  &  -1.946126    &   -1.927259   &   -1.914140   &   -1.906981   \\
        & -2.034361&  -2.073209  &    -2.116190  &    -2.162958  &    -2.213182  \\
1.0& -1.969694  &  -1.945683    &   -1.928286    &  -1.917745    &  -1.914214    \\
        & -2.036229 & -2.077973    &   -2.124808   &  -2.176305    &  -2.232051    \\
&&&&&\\
        &0.6&           0.7&            0.8&            0.9&            1.0\\
        \hline
0.0& -1.890016  &   -1.882792   &  -1.879170    &  -1.879284   &  -1.883249    \\
      & -2.198388 &    -2.237750  &   -2.278567    &  -2.320724   &  -2.364112     \\
0.2& -1.888950  &   -1.883170   &  -1.881506    &  -1.884089   &  -1.891014    \\
      & -2.210546 &    -2.252476  &   -2.295943   &   -2.340814   &  -2.386963     \\
0.4& -1.891353  &   -1.888185   &  -1.889758    &  -1.896163   &  -1.907439    \\
      & -2.225963 &    -2.271567  &   -2.318931   &   -2.367895   &  -2.418308    \\
0.6& -1.897068  &   -1.897568   &  -1.903499    &  -1.914877   &  -1.931650    \\
      & -2.244637 &    -2.295011  &   -2.347498   &   -2.401899   &  -2.458034   \\
0.8& -1.905924 &    -1.911032  &   -1.922289   &  -1.939595    &  -1.962780    \\
      & -2.266546 &    -2.322758  &   -2.381549    & -2.442671    & -2.505900     \\
1.0& -1.917745  &   -1.928286   &  -1.945683    & -1.969694    &  -2.000000    \\
      & -2.291647 &    -2.354724  &  -2.420937    &  -2.489975    &  -2.561553    \\
\hline
\end{tabular}
\caption{Eigenvalues for $j = 3/2, m = \pm 1/2$ and $P = -1$.}
\label{table-j3-m1-P-1}
\end{table}

\begin{table}
\begin{tabular}{|c|ccccc|}
\hline
&\multicolumn{5}{c|}{k} \\
n&0&1&2&3&4\\ \hline
1&$-1$&$3.33287\times10^{-1}$&$1.66774\times10^{-4}$&$-1.82477\times10^{-4}$&$2.69509\times10^{-4}$\\
&1&-0.333356&$7.8727\times10^{-5}$&$4.68693\times10^{-5}$&$-7.76405\times10^{-5}$ \\
2&0&$2.90019\times10^{-4}$&$7.32430\times10^{-2}$&$7.62158\times10^{-4}$&$-2.00106\times10^{-3}$\\
&0&$8.39768\times10^{-5}$&0.0737528&$-7.84339\times10^{-4}$&$8.33219\times10^{-4}$ \\
3&0&$-3.56377\times10^{-4}$&$2.97546\times10^{-2}$&$-5.99395\times10^{-3}$&$4.35512\times10^{-3}$\\
&0&$-1.55588\times10^{-4}$&-0.0288713&$8.79491\times10^{-3}$&$-2.42401\times10^{-3}$ \\
4&0&$-1.87177\times10^{-5}$&$1.61106\times10^{-3}$&$4.59520\times10^{-3}$&$-8.23406\times10^{-3}$\\
&0&$1.33331\times10^{-4}$&$-8.50242\times10^{-4}$&$4.21954\times10^{-3}$&$-1.17962\times10^{-3}$ \\ \hline
\end{tabular}
\caption{Fitting coefficients for $j = 1/2, m = \pm 1/2$ and $P = +1$. The upper line in each row is for $m=1/2$, and the lower line for $m=-1/2$.}
\label{fit-j1-P1}
\end{table}

\begin{table}
\begin{tabular}{|c|ccccc|}
\hline
&\multicolumn{5}{c|}{k} \\
n&0&1&2&3&4\\ \hline
1&$3.33047\times10^{-1}$&$3.31222\times10^{-1}$&$3.98816\times10^{-3}$&$-2.03543\times10^{-3}$&$2.37639\times10^{-4}$\\
&$-3.32884\times10^{-1}$&$-3.34729\times10^{-1}$&$1.82192\times10^{-3}$&$-1.36801\times10^{-3}$&$4.66671\times10^{-4}$\\
2&$-2.92723\times10^{-1}$&$3.12502\times10^{-1}$&$-1.08338\times10^{-1}$&$1.89525\times10^{-2}$&$-2.68764\times10^{-3}$\\
&$-3.00825\times10^{-1}$&$3.09671\times10^{-1}$&$-8.99404\times10^{-2}$&$1.03932\times10^{-2}$&$-3.18256\times10^{-3}$\\
3&$-8.06550\times10^{-2}$&$6.85237\times10^{-3}$&$9.79197\times10^{-2}$&$-6.14738\times10^{-2}$&$9.59822\times10^{-3}$\\
&$8.14284\times10^{-2}$&$-8.29826\times10^{-2}$&$3.77385\times10^{-2}$&$-2.10873\times10^{-2}$&$7.55711\times10^{-3}$\\
4&$4.03322\times10^{-2}$&$-5.06553\times10^{-2}$&$-6.96088\times10^{-3}$&$2.74214\times10^{-2}$&$-8.09157\times10^{-3}$\\
&$-9.26754\times10^{-3}$&$4.99333\times10^{-3}$&$2.46876\times10^{-3}$&$1.95452\times10^{-3}$&$-2.47130\times10^{-3}$\\
\hline
\end{tabular}
\caption{Fitting coefficients for $j = 1/2, m = \pm 1/2$ and $P = -1$. The upper line in each row is for $m=1/2$, and the lower line for $m=-1/2$.}
\label{fit-j1-P-1}
\end{table}

\begin{table}
\begin{tabular}{|c|ccccc|}
\hline
&\multicolumn{5}{c|}{k} \\
n&0&1&2&3&4\\ \hline
1&-1&$1.99945\times10^{-1}$&$3.71992\times10^{-4}$&$-7.87532\times10^{-4}$&$4.65803\times10^{-4}$\\
&1&$-2.00063\times10^{-1}$&$2.84647\times10^{-4}$&$-3.91985\times10^{-4}$&$1.89222\times10^{-4}$\\
2&0&$2.87850\times10^{-4}$&$2.99573\times10^{-2}$&$4.57896\times10^{-3}$&$-2.73991\times10^{-3}$\\
&0&$3.72905\times10^{-4}$&$3.02612\times10^{-2}$&$2.33862\times10^{-3}$&$-1.13402\times10^{-3}$\\
3&0&$-4.81935\times10^{-4}$&$1.25710\times10^{-2}$&$-8.94371\times10^{-3}$&$4.94845\times10^{-3}$\\
&0&$-6.35154\times10^{-4}$&$-6.12322\times10^{-3}$&$-3.03913\times10^{-3}$&$1.83562\times10^{-3}$\\
4&0&$2.62158\times10^{-4}$&$-1.84259\times10^{-3}$&$7.31973\times10^{-3}$&$-3.94239\times10^{-3}$\\
&0&$3.25972\times10^{-4}$&$-1.57711\times10^{-3}$&$3.59247\times10^{-3}$&$-1.49975\times10^{-3}$\\
\hline
\end{tabular}
\caption{Fitting coefficients for $j = 3/2, m = \pm 3/2$ and $P = +1$. The upper line in each row is for $m=3/2$, and the lower line for $m=-3/2$.}
\label{fit-j3-P1}
\end{table}

\begin{table}
\begin{tabular}{|c|ccccc|}
\hline
&\multicolumn{5}{c|}{k} \\
n&0&1&2&3&4\\ \hline
1&$5.99797\times10^{-1}$&$2.00423\times10^{-1}$&$3.40583\times10^{-5}$&$-4.77152\times10^{-4}$&$2.28612\times10^{-4}$\\
&$-6.00024\times10^{-1}$&$-2.00033\times10^{-1}$&$2.87207\times10^{-5}$&$7.14085\times10^{-5}$&$-6.24219\times10^{-5}$\\
2&$-1.26255\times10^{-1}$&$1.24048\times10^{-1}$&$-3.12219\times10^{-2}$&$2.82263\times10^{-3}$&$-1.47819\times10^{-3}$\\
&$-1.27827\times10^{-1}$&$1.28410\times10^{-1}$&$-3.22650\times10^{-2}$&$-8.40480\times10^{-4}$&$6.80416\times10^{-4}$\\
3&$-3.53405\times10^{-2}$&$3.91967\times10^{-2}$&$-9.37469\times10^{-3}$&$-5.64216\times10^{-3}$&$3.06728\times10^{-3}$\\
&$3.03746\times10^{-2}$&$-2.96270\times10^{-2}$&$6.62180\times10^{-3}$&$2.07091\times10^{-3}$&$-1.47298\times10^{-3}$\\
4&$2.41821\times10^{-4}$&$-1.29284\times10^{-2}$&$1.51332\times10^{-2}$&$-3.26087\times10^{-3}$&$-9.82596\times10^{-4}$\\
&$-4.08651\times10^{-3}$&$3.00329\times10^{-3}$&$2.36225\times10^{-3}$&$-3.51752\times10^{-3}$&$1.39410\times10^{-3}$\\
\hline
\end{tabular}
\caption{Fitting coefficients for $j = 3/2, m = \pm 3/2$ and $P = -1$. The upper line in each row is for $m=3/2$, and the lower line for $m=-3/2$.}
\label{fit-j3-P-1}
\end{table}

\begin{table}
\begin{tabular}{|c|ccccc|}
\hline
&\multicolumn{5}{c|}{k} \\
n&0&1&2&3&4\\ \hline
1&$-3.33047\times10^{-1}$&$6.87003\times10^{-2}$&$-3.52549\times10^{-3}$&$1.25094\times10^{-3}$&$1.90819\times10^{-4}$\\
&$3.32884\times10^{-1}$&$-6.52992\times10^{-2}$&$-1.56447\times10^{-3}$&$8.64220\times10^{-4}$&$-2.19611\times10^{-4}$\\
2&$2.92722\times10^{-1}$&$-3.12089\times10^{-1}$&$1.53920\times10^{-1}$&$-1.50603\times10^{-2}$&$5.51959\times10^{-4}$\\
&$3.00824\times10^{-1}$&$-3.09458\times10^{-1}$&$1.36164\times10^{-1}$&$-6.81511\times10^{-3}$&$1.38445\times10^{-3}$\\
3&$8.06564\times10^{-2}$&$-7.56215\times10^{-3}$&$-8.92653\times10^{-2}$&$5.52054\times10^{-2}$&$-6.37204\times10^{-3}$\\
&$-8.14264\times10^{-2}$&$8.25932\times10^{-2}$&$-3.92363\times10^{-2}$&$1.47876\times10^{-2}$&$-4.17156\times10^{-3}$\\
4&$-4.03332\times10^{-2}$&$5.10444\times10^{-2}$&$4.71906\times10^{-3}$&$-2.46746\times10^{-2}$&$6.22122\times10^{-3}$\\
&$9.26669\times10^{-3}$&$-4.80496\times10^{-3}$&$-3.88778\times10^{-3}$&$2.34438\times10^{-3}$&$-1.17466\times10^{-4}$\\
\hline
\end{tabular}
\caption{Fitting coefficients for $j = 3/2, m = \pm 1/2$ and $P = +1$. The upper line in each row is for $m=1/2$, and the lower line for $m=-1/2$.}
\label{fit-j3-m1-P1}
\end{table}

\begin{table}
\begin{tabular}{|c|ccccc|}
\hline
&\multicolumn{5}{c|}{k} \\
n&0&1&2&3&4\\ \hline
1&$1.99974\times10^{-1}$&$6.65908\times10^{-2}$&$2.08259\times10^{-4}$&$-2.82760\times10^{-4}$&$-5.93627\times10^{-5}$\\
&$-1.99866\times10^{-1}$&$-6.69306\times10^{-2}$&$9.28649\times10^{-5}$&$1.25203\times10^{-4}$&$-8.67219\times10^{-5}$\\
2&$-1.91707\times10^{-1}$&$1.92687\times10^{-1}$&$-1.24489\times10^{-1}$&$3.25586\times10^{-3}$&$2.08439\times10^{-4}$\\
&$-1.93239\times10^{-1}$&$1.94559\times10^{-1}$&$-1.23591\times10^{-1}$&$-8.37842\times10^{-5}$&$2.56220\times10^{-4}$\\
3&$-1.65338\times10^{-2}$&$1.26573\times10^{-2}$&$-2.56257\times10^{-2}$&$-2.31309\times10^{-3}$&$-8.48344\times10^{-4}$\\
&$1.90423\times10^{-2}$&$-2.18047\times10^{-2}$&$3.85236\times10^{-2}$&$-9.44174\times10^{-3}$&$1.13187\times10^{-3}$\\
4&$8.26685\times10^{-3}$&$-1.46583\times10^{-2}$&$7.67572\times10^{-3}$&$-5.46338\times10^{-3}$&$7.20356\times10^{-3}$\\
&$1.78362\times10^{-3}$&$-4.28852\times10^{-3}$&$2.04973\times10^{-3}$&$-3.65754\times10^{-3}$&$1.31285\times10^{-3}$\\
\hline
\end{tabular}
\caption{Fitting coefficients for $j = 3/2, m = \pm 1/2$ and $P = -1$. The upper line in each row is for $m=1/2$, and the lower line for $m=-1/2$.}
\label{fit-j3-m1-P-1}
\end{table}

\end{document}